\documentclass[12pt]{article}
\usepackage{a4}
\usepackage{epsfig}
\usepackage{latexsym}

\newtheorem{theorem}{Theorem}
\newtheorem{lemma}{Lemma}
\newtheorem{corollary}{Corollary}
\newtheorem{definition}{Definition}

\newtheorem{invariant}{Invariant}
\newtheorem{datastructure}{Data Structure}

\newcommand{\fullpaper}[1]{}
\newcommand{\reducepar}{\vspace{-3.5mm}}

\newcommand{\define}[1]{{\sc #1}}

\newcommand{\gimmebreak}{\medskip\centerline{$\triangleleft\diamond\triangleright$}\medskip}

\newenvironment{proof}
{\noindent   {\bf Proof.}}{\hspace*{\fill}$\Box$\par\vspace{2mm}}

\newenvironment{mylist}[1]{
\setbox1=\hbox{#1}
\begin{list}{}{
\setlength{\labelwidth}{\wd1}
\setlength{\leftmargin}{\wd1}
\addtolength{\leftmargin}{1em}
\addtolength{\leftmargin}{\labelsep}
\setlength{\rightmargin}{1em}}}{\end{list}}

\newcommand{\litem}[1]{\item[#1\hfill]}

\newcommand{\LaBeL}{UNO}
\newcommand{\CaPtIoN}{DUE}

\newcommand{\FoNtSiZe}{ONE}
\newenvironment{frameprog}[2]{%
\renewcommand{\FoNtSiZe}{#2}%
\begin{minipage}{#1}%
\begin{prog}{\FoNtSiZe}%
}{\end{prog}%
\vspace{-.15in}%
\end{minipage}%
}

\newcounter{stepcount}

\newcounter{poi}
\newcommand{\N}{\stepcounter{poi} \< \thepoi . \>}

\newenvironment{prog}[1]{#1 
\setcounter{poi}{0}
\vspace{-.15in}
\begin{tabbing}
=spa\=spa\=spa\=spa\=spa\=spa\=spa\=spa\=spa\=spa\=spa\=spa\=spa\=\kill
\+\\}{
\end{tabbing}
\vspace{-.15in}
}

\newcommand {\BEGIN}{\mbox{\bf begin}}
\newcommand {\DO}{\mbox{\bf do}}

\newcommand {\ELSE}{\mbox{\bf else}}
\newcommand {\END}{\mbox{\bf end}}

\newcommand {\FOR}{\mbox{\bf for}}
\newcommand {\FOREACH}{\mbox{\bf for each}}
\newcommand {\FUNCTION}{\mbox{\bf function}}
\newcommand {\IF}{\mbox{\bf if}}

\newcommand {\NOT}{\mbox{\bf not}}

\newcommand {\PROCEDURE}{\mbox{\bf procedure}}

\newcommand {\RETURN}{\mbox{\bf return}}

\newcommand {\THEN}{\mbox{\bf then}}
\newcommand {\TO}{\mbox{\bf to}}

\begin{document}

\title{{\Large\bf Mantaining Dynamic Matrices \\ for Fully Dynamic 
Transitive Closure}~\thanks{This work has been partially supported by 
the IST Programme of the EU under contract n.~IST-1999-14.186 
(ALCOM-FT), by the Italian Ministry of University and Scientific 
Research (Project ``Algorithms for Large Data Sets: Science and 
Engineering'' and by CNR, the Italian National Research Council under 
contract n.~00.00346.CT26.  This work is based on the first author's 
PhD Thesis~\cite{D01} and a preliminary version has been presented at 
the {\em 41st Annual Symp.~on Foundations of Computer Science\/} (FOCS 
2000)~\cite{DI00}.  }}

\author{ {\em Camil Demetrescu}~\thanks{Email: {\tt 
demetres@dis.uniroma1.it}.  URL: {\tt 
http://www.dis.uniroma1.it/\symbol{126}demetres}.  Part of this work 
has been done while visiting AT\&T Shannon Laboratory, Florham Park, 
NJ.}\\
Dipartimento di Informatica e Sistemistica\\
Universit\`a di Roma ``La Sapienza'', Roma, Italy \and {\em Giuseppe 
F. Italiano}~\thanks{Email: {\tt italiano@info.uniroma2.it}.  URL: 
{\tt http://www.info.uniroma2.it/\symbol{126}italiano}.  Part of this 
work has been done while visiting Columbia University, New York, 
NY.}\\
Dipartimento di Informatica, Sistemi e Produzione\\
Universit\`a di Roma ``Tor Vergata'', Roma, Italy
}

\date{}

\maketitle

\begin{abstract}
In this paper we introduce a general framework for casting fully
dynamic transitive closure into the problem of reevaluating
polynomials over matrices.  With this technique, we improve the best
known bounds for fully dynamic transitive closure.  In particular, we
devise a deterministic algorithm for general directed graphs that
achieves $O(n^2)$ amortized time for updates, while preserving unit
worst-case cost for queries.  In case of deletions only, our algorithm
performs updates faster in $O(n)$ amortized time.

Our matrix-based approach yields an algorithm for directed acyclic 
graphs that breaks through the $O(n^2)$ barrier on the 
single-operation complexity of fully dynamic transitive closure.  We 
can answer queries in $O(n^\epsilon)$ time and perform updates in 
$O(n^{\omega(1,\epsilon,1)-\epsilon}+n^{1+\epsilon})$ time, for any 
$\epsilon\in[0,1]$, where $\omega(1,\epsilon,1)$ is the exponent of 
the multiplication of an $n\times n^{\epsilon}$ matrix by an 
$n^{\epsilon}\times n$ matrix.  The current best bounds on 
$\omega(1,\epsilon,1)$ imply an $O(n^{0.58})$ query time and an 
$O(n^{1.58})$ update time.  Our subquadratic algorithm is randomized, 
and has one-side error.
\end{abstract}

\newpage

\section{Introduction}
\label{se:intro}

In this paper we present fully dynamic algorithms for maintaining the
transitive closure of a directed graph.  A dynamic graph algorithm
maintains a given property on a graph subject to dynamic changes, such
as edge insertions and edge deletions.  We say that an algorithm is
{\em fully dynamic} if it can handle both edge insertions and edge
deletions.  A {\em partially dynamic} algorithm can handle either edge
insertions or edge deletions, but not both: we say that it is {\em
incremental} if it supports insertions only, and {\em decremental} if
it supports deletions only.  In the {\em fully dynamic transitive
closure problem} we wish to maintain a directed graph $G=(V,E)$ under
an intermixed sequence of the following operations:

\begin{mylist}{{\em Insert$(x,y)$}:   }

\litem{{\em Insert$(x,y)$}:} insert an edge from $x$ to
$y$ in $G$;

\litem{{\em Delete$(x,y)$}:} delete the edge from $x$ to
$y$ in $G$;

\litem{{\em Query$(x,y)$}:} report {\em yes} if there is a
path from $x$ to $y$ in $G$, and {\em no} otherwise.

\end{mylist}

\noindent Throughout the paper, we denote by $m$ and by $n$ the number
of edges and vertices in $G$, respectively.

Research on dynamic transitive closure spans over two decades.  Before 
describing the results known, we list the bounds obtainable with 
simple-minded methods.  If we do nothing during each update, then we 
have to explore the whole graph in order to answer reachability 
queries: this gives $O(n^2)$ time per query and $O(1)$ time per update 
in the worst case.  On the other extreme, we could recompute the 
transitive closure from scratch after each update; as this task can be 
accomplished via matrix multiplication~\cite{AHU74,M71}, this approach 
yields $O(1)$ time per query and $O(n^\omega)$ time per update in the 
worst case, where $\omega$ is the best known exponent for matrix 
multiplication (currently $\omega<2.38$~\cite{CW90}).

\paragraph{Previous Work.}

For the {\em incremental} version of the problem, the first algorithm 
was proposed by Ibaraki and Katoh~\cite{IK83} in 1983: its running 
time was $O(n^3)$ over any sequence of insertions.  This bound was 
later improved to $O(n)$ amortized time per insertion by 
Italiano~\cite{I86} and also by La Poutr\'e and van 
Leeuwen~\cite{LvL88}.  Yellin~\cite{Y93} gave an $O(m^*\delta_{max})$ 
algorithm for $m$ edge insertions, where $m^*$ is the number of edges 
in the final transitive closure and $\delta_{max}$ is the maximum 
out-degree of the final graph.  All these algorithms maintain 
explicitly the transitive closure, and so their query time is $O(1)$.

The first {\em decremental} algorithm was again given by Ibaraki and 
Katoh~\cite{IK83}, with a running time of $O(n^2)$ per deletion.  This 
was improved to $O(m)$ per deletion by La Poutr\'e and van 
Leeuwen~\cite{LvL88}.  Italiano~\cite{I88} presented an algorithm that 
achieves $O(n)$ amortized time per deletion on directed acyclic 
graphs.  Yellin~\cite{Y93} gave an $O(m^*\delta_{max})$ algorithm for 
$m$ edge deletions, where $m^*$ is the initial number of edges in the 
transitive closure and $\delta_{max}$ is the maximum out-degree of the 
initial graph.  Again, the query time of all these algorithms is 
$O(1)$.  More recently, Henzinger and King~\cite{HK95} gave a 
randomized decremental transitive closure algorithm for general 
directed graphs with a query time of $O(n/\log n)$ and an amortized 
update time of $O(n\log^2 n)$.

The first {\em fully dynamic} transitive closure algorithm was devised
by Henzinger and King~\cite{HK95} in 1995: they gave a randomized
Monte Carlo algorithm with one-side error supporting a query time of
$O(n/\log n)$ and an amortized update time of $O(n\hat{m}^{0.58}\log^2
n)$, where $\hat{m}$ is the average number of edges in the graph
throughout the whole update sequence.  Since $\hat{m}$ can be as high
as $O(n^2)$, their update time is $O(n^{2.16}\log^2 n)$.  Khanna,
Motwani and Wilson~\cite{KMW96} proved that, when a lookahead of
$\Theta(n^{0.18})$ in the updates is permitted, a deterministic update
bound of $O(n^{2.18})$ can be achieved.  Very recently, King and
Sagert~\cite{KS99} showed how to support queries in $O(1)$ time and
updates in $O(n^{2.26})$ time for general directed graphs and $O(n^2)$
time for directed acyclic graphs; their algorithm is randomized with
one-side error.  The bounds of King and Sagert were further improved
by King~\cite{K99}, who exhibited a deterministic algorithm on general
digraphs with $O(1)$ query time and $O(n^2\log n)$ amortized time per
update operations, where updates are insertions of a set of edges
incident to the same vertex and deletions of an arbitrary subset of
edges.  We remark that all these algorithms (except~\cite{K99}) use
fast matrix multiplication as a subroutine.

We observe that fully dynamic transitive closure algorithms with
$O(1)$ query time maintain explicitly the transitive closure of the
input graph, in order to answer each query with exactly one lookup (on
its adjacency matrix).  Since an update may change as many as $\Omega
(n^2)$ entries of this matrix, $O(n^2)$ seems to be the best update
bound that one could hope for this class of algorithms.  It is thus
quite natural to ask whether the $O(n^2)$ update bound can be actually
realized for fully dynamic transitive closure on general directed
graphs while maintaining one lookup per query.  \fullpaper{It is
remarkable that designing such an algorithm has been an elusive goal
for many years, while the algorithm with unit-cost update and $O(n^2)$
query is rather trivial.}Another important question, if one is willing
to spend more time for queries, is whether the $O(n^2)$ barrier for
the single-operation time complexity of fully dynamic transitive
closure can be broken.  We remark that this has been an elusive goal
for many years.

\reducepar
\paragraph{Our Results.}
In this paper, we affirmatively answer both questions.  We first 
exhibit a deterministic algorithm for fully dynamic transitive closure 
on general digraphs that does exactly one matrix look-up per query and 
supports updates in $O(n^2)$ amortized time, thus improving 
over~\cite{K99}.  Our algorithm can also support within the same time 
bounds the generalized updates of~\cite{K99}, i.e., insertion of a set 
of edges incident to the same vertex and deletion of an arbitrary 
subset of edges.  In the special case of deletions only, our algorithm 
achieves $O(n)$ amortized time for deletions and $O(1)$ time for 
queries: this generalizes to directed graphs the bounds of~\cite{I88}, 
and improves over~\cite{HK95}.

As our second contribution, we present the first algorithm that breaks 
through the $O(n^2)$ barrier on the single-operation time complexity 
of fully dynamic transitive closure.  In particular, we show how to 
trade off query times for updates on directed acyclic graphs: each 
query can be answered in time $O(n^\epsilon)$ and each update can be 
performed in time 
$O(n^{\omega(1,\epsilon,1)-\epsilon}+n^{1+\epsilon})$, for any 
$\epsilon\in[0,1]$, where $\omega(1,\epsilon,1)$ is the exponent of 
the multiplication of an $n\times n^{\epsilon}$ matrix by an 
$n^{\epsilon}\times n$ matrix.  Balancing the two terms in the update 
bound yields that $\epsilon$ must satisfy the equation 
$\omega(1,\epsilon,1)=1+2\epsilon$.  The current best bounds on 
$\omega(1,\epsilon,1)$~\cite{CW90,HP98} imply that 
$\epsilon<0.58$~\cite{Z98}.  Thus, the smallest update time is 
$O(n^{1.58})$, which gives a query time of $O(n^{0.58})$.  Our 
subquadratic algorithm is randomized, and has one-side error.

All our algorithms are based on a novel technique: we introduce a
general framework for maintaining polynomials defined over matrices,
and we cast fully dynamic transitive closure into this framework.  In
particular, our deterministic algorithm hinges upon the equivalence
between transitive closure and matrix multiplication on a closed
semiring; this relation has been known for over $30$ years (see e.g.,
the results of Munro \cite{M71}, Furman \cite{F70} and Fischer and
Meyer~\cite{FM71}) and yields the fastest known static algorithm for
transitive closure.  Surprisingly, no one before seems to have
exploited this equivalence in the dynamic setting: some recent
algorithms \cite{HK95,KMW96,KS99} make use of fast matrix
multiplication, but only as a subroutine for fast updates.
Differently from other approaches, the crux of our method is to use
dynamic reevaluation of products of Boolean matrices as the kernel for
solving dynamic transitive closure.

The remainder of this paper is organized as follows.  We first 
formally define the fully dynamic transitive closure problem and we 
give preliminary definitions in Section~\ref{se:tc-defs}.  A 
high-level overview of our approach is given in 
Section~\ref{se:tc-overview}.  In Section~\ref{se:tc-matrices} we 
introduce two problems on dynamic matrices, and show how to solve them 
efficiently.  Next, we show how to exploit these problems on dynamic 
matrices for the design of three efficient fully dynamic algorithms 
for transitive closure in Section~\ref{se:tc-log}, 
Section~\ref{se:tc-divcon} and Section~\ref{se:tc-subquad}, 
respectively.  Finally, in Section~\ref{se:tc-conclusions} we list 
some concluding remarks.

\section{Fully Dynamic Transitive Closure}
\label{se:tc-defs}

In this section we give a more formal definition of the fully dynamic 
transitive closure problem considered in this paper.  We assume the 
reader to be familiar with standard graph and algebraic terminology as 
contained for instance in~\cite{AHU74,CLR90}.

\begin{figure}
\centerline{ { \epsfxsize=15cm \epsffile{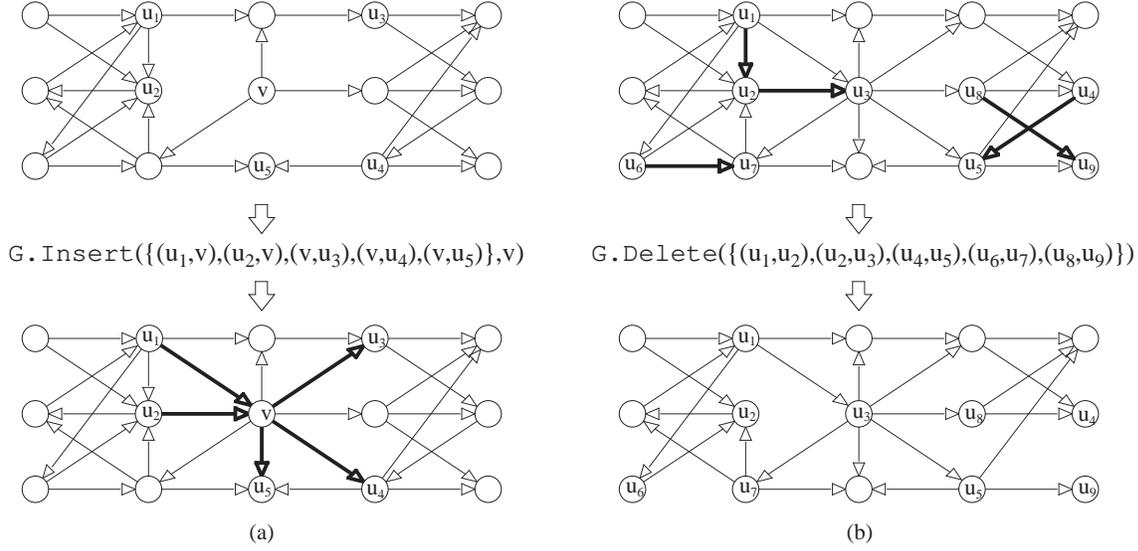}} }
\caption{(a) {\tt Insert} operation; (b) {\tt Delete}
operation as in Definition~\ref{def:fdtc}.}
\label{fi:tc-operations}
\end{figure}

\begin{definition}
\label{def:fdtc}
Let $G=(V,E)$ be a directed graph and let $TC(G)=(V,E')$ be its 
transitive closure.  The \define{Fully Dynamic Transitive Closure 
Problem} consists of maintaining a data structure {\tt G} for graph 
$G$ under an intermixed sequence $\sigma=\langle {\tt 
G.Op}_1,\ldots,{\tt G.Op}_k\rangle$ of \define{Initialization}, 
\define{Update}, and \define{Query} operations.  Each operation ${\tt 
G.Op}_j$ on data structure {\tt G} can be either one of the following:

\begin{itemize}

\item {\tt G.Init}$(A)$: perform the initialization operation
$E\leftarrow A$, where $A\subseteq V\times V$.

\item {\tt G.Insert}$(v,I)$: perform the update $E\leftarrow 
E\cup\{(u,v)~|~u\in V \wedge (u,v)\in I\} \cup\{(v,u)~|~u\in V \wedge 
(v,u)\in I\}$, where $I\subseteq E$ and $v\in V$.  We call this update 
a $v$-\define{Centered} insertion in $G$.

\item {\tt G.Delete}$(D)$: perform the update $E\leftarrow E-D$, where
$D\subseteq E$.

\item {\tt G.Query}$(x,y)$: perform a query operation on $TC(G)$ by
returning $1$ if $(x,y)\in E'$ and $0$ otherwise.

\end{itemize}

\end{definition}

Few remarks are in order at this point.  First, the generalized {\tt 
Insert} and {\tt Delete} updates considered here have been first 
introduced by King in~\cite{K99}.  With just one operation, they are 
able to change the graph by adding or removing a whole set of edges, 
rather than a single edge, as illustrated in 
Figure~\ref{fi:tc-operations}.  \fullpaper{Notice that we provide an 
operational (and not algorithmic) definition of operations, giving no 
detail about what the actual implementation should do for supporting 
them.} Second, we consider explicitly initializations of the graph $G$ 
and, more generally than in the traditional definitions of dynamic 
problems, we allow them to appear everywhere in sequence $\sigma$.  
This gives more generality to the problem, and allows for more 
powerful data structures, i.e., data structures that can be restarted 
at run time on a completely different input graph.  Differently from 
others variants of the problem, we do not address the issue of 
returning actual paths between nodes, and we just consider the problem 
of answering reachability queries.

\gimmebreak

It is well known that, if $G=(V,E)$ is a directed graph and $X_G$ is 
its adjacency matrix, computing the Kleene closure $X_G^*$ of $X_G$ is 
equivalent to computing the (reflexive) transitive closure $TC(G)$ of 
$G$.  For this reason, in this paper, instead of considering directly 
the problem introduced in Definition~\ref{def:fdtc}, we study an 
equivalent problem on matrices.  Before defining it formally, we need 
some preliminary notation.

\begin{definition}
If $X$ is a matrix, we denote by $I_{X,i}$ and $J_{X,j}$ the matrices
equal to $X$ in the $i$-th row and $j$-th column, respectively, and
null in any other entries:

$$
I_{X,i}[x,y]=
\left\{
\begin{array}{ll}
    X[x,y] & if~x=i      \\
    0 & \mbox{otherwise} \\
\end{array}
\right.
$$

$$
J_{X,i}[x,y]=
\left\{
\begin{array}{ll}
    X[x,y] & if~y=i      \\
    0 & \mbox{otherwise} \\
\end{array}
\right.
$$

\end{definition}

\begin{definition}
Let $X$ and $Y$ be $n\times n$ Boolean matrices.  Then $X\subseteq Y$
if and only if $X[x,y]=1~\Rightarrow~Y[x,y]=1$ for any $x,y\in
\{1,\ldots,n\}$.
\end{definition}

We are now ready to define a dynamic version of the problem of 
computing the Kleene closure of a Boolean matrix.  In what follows, we 
assume that algebraic operations $+$ and $-$ are performed modulo 
$n+1$ by looking at Boolean values $0$ and $1$ as integer numbers.  
Integer results are binarized by converting back nonzero values into 
$1$ and zero values into $0$.  We remark that in our dynamic setting 
operator $-$ is just required to flip matrix entries from $1$ to $0$.

\begin{definition}
\label{def:fdkc}
Let $X$ be an $n\times n$ Boolean matrix and let $X^*$ be its Kleene 
closure.  We define the \define{Fully Dynamic Boolean Matrix Closure 
Problem} as the problem of maintaining a data structure {\tt X} for 
matrix $X$ under an intermixed sequence $\sigma=\langle {\tt 
X.Op}_1,\ldots,{\tt X.Op}_k\rangle$ of initialization, update, and 
query operations.  Each operation ${\tt X.Op}_j$ on data structure 
{\tt X} can be either one of the following:

\begin{itemize}

\item {\tt X.Init}$^*(Y)$: perform the initialization operation
$X\leftarrow Y$, where $Y$ is an $n\times n$ Boolean matrix.

\item {\tt X.Set}$^*(i,\Delta X)$: perform the update $X\leftarrow 
X+I_{\Delta X,i}+J_{\Delta X,i}$, where $\Delta X$ is an $n\times n$ 
Boolean matrix and $i\in \{1,\ldots,n\}$.  We call this kind of update 
an $i$-\define{Centered} set operation on $X$ and we call $\Delta X$ 
\define{Update Matrix}.

\item {\tt X.Reset}$^*(\Delta X)$: perform the update $X\leftarrow
X-\Delta X$, where $\Delta X\subseteq X$ is an $n\times n$ Boolean
update matrix.

\item {\tt X.Lookup}$^*(x,y)$: return the value of $X^*[x,y]$, where
$x,y\in \{1,\ldots,n\}$.

\end{itemize}

\end{definition}

Notice that {\tt Set}$^*$ is allowed to modify only the $i$-th row and 
the $i$-th column of $X$, while {\tt Reset}$^*$ and {\tt Init}$^*$ can 
modify any entries of $X$.  We stress the strong correlation between 
Definition~\ref{def:fdkc} and Definition~\ref{def:fdtc}: if $G$ is a 
graph and $X$ is its adjacency matrix, operations {\tt X.Init}$^*$, 
{\tt X.Set}$^*$, {\tt X.Reset}$^*$, and {\tt X.Lookup}$^*$ are 
equivalent to operations {\tt G.Init}, {\tt G.Insert}, {\tt G.Delete}, 
and {\tt G.Query}, respectively.

\section{Overview of Our Approach}
\label{se:tc-overview}

In this section we give an overview of the new ideas presented in this 
paper, discussing the most significant aspects of our techniques.

Our approach consists of reducing fully dynamic transitive closure to 
the problem of maintaining efficiently polynomials over matrices 
subject to updates of their variables.  In particular, we focus on the 
equivalent problem of fully dynamic Kleene closure and we show that 
efficient data structures for it can be realized using efficient data 
structures for maintaining polynomials over matrices.

Suppose that we have a polynomial over Boolean matrices, e.g., 
$P(X,Y,Z,W)=X+YZ^2W$, where matrices $X$, $Y$, $Z$ and $W$ are its 
variables.  The value $P(X,Y,Z,W)$ of the polynomial can be computed 
via sum and multiplication of matrices $X$, $Y$, $Z$ and $W$ in 
$O(n^{2.38})$.  Now, what kind of modifications can we perform on a 
variable, e.g., variable $Z$, so as to have the chance of updating the 
value of $P(X,Y,Z,W)$ in less than $O(n^{2.38})$ time?

In Section~\ref{se:tc-polynomials} we show a data structure that 
allows us to reevaluate correctly $P(X,Y,Z,W)$ in just $O(n^2)$ 
amortized time after flipping to $1$ any entries of $Z$ that were 
$0$, provided they lie on a row or on a column ({\tt SetRow} or {\tt 
SetCol} operation), of after flipping to $0$ {\em any} entries of $Z$ 
that were $1$ ({\tt Reset} operation).  This seems a step forward, 
but are this kind of updates of variables powerful enough to be useful 
our original problem of fully dynamic transitive closure?  
Unfortunately, the answer is no.  Actually, we also require the more 
general {\tt Set} operation of flipping to $1$ {\em any} entries of 
$Z$ that were $0$.  Now, if we want to have our polynomial always up 
to date after each variable change of this kind, it seems that there 
is no way of doing any better than recomputing everything from 
scratch.

So let us lower our expectations on our data structure for maintaining 
$P$, and tolerate errors.  In exchange, our data structure must 
support efficiently the general {\tt Set} operation.  The term 
``errors'' here means that we maintain a ``relaxed'' version of the 
correct value of the polynomial, where some $0$'s may be incorrect.  
The only important property that we require is that any $1$'s that 
appear in the correct value of the polynomial after performing a {\tt 
SetRow} or {\tt SetCol} operation must also appear in the relaxed 
value that we maintain.  This allows us to support any {\tt Set} 
operation efficiently in a lazy fashion (so in the following we call 
it {\tt LazySet}) and is powerful enough for our original problem of 
fully dynamic transitive closure.

Actually, doing things lazily while maintaining the desired properties 
in our data structure for polynomials is the major technical 
difficulty in Section~\ref{se:tc-polynomials}.  Sections 
\ref{se:tc-log} and \ref{se:tc-divcon} then show two methods to solve 
the fully dynamic Boolean matrix closure problem by using polynomials 
of Boolean matrices as if they were building blocks.  The second 
method yields the fastest known algorithm for fully dynamic transitive 
closure with constant query time.  If we give up maintaining 
polynomials of degree $>1$, using a surprisingly simple lazy technique 
we can even support certain kinds of variable updates in subquadratic 
worst-case time per operation (see 
Section~\ref{se:tc-subquadraticmatrices}).  This turns out to be once 
again applicable to fully dynamic transitive closure, yielding the 
first subquadratic algorithms known so far for the problem (see 
Section~\ref{se:tc-subquad}).

\section{Dynamic Matrices}
\label{se:tc-matrices}

In this section we consider two problems on dynamic matrices and we 
devise fast algorithms for solving them.  As we already stated, these 
problems will be central to designing efficient algorithms for the 
fully dynamic Boolean matrix closure problem introduced in 
Definition~\ref{def:fdkc}.  In more detail, in 
Section~\ref{se:tc-polynomials} we address the problem of reevaluating 
polynomials over Boolean matrices under modifications of their 
variables.  We propose a data structure for maintaining efficiently 
the special class of polynomials of degree $2$ consisting of single 
products of Boolean matrices.  We show then how to use this data 
structure for solving the more general problem on arbitrary 
polynomials.  In Section~\ref{se:tc-subquadraticmatrices} we study the 
problem of finding an implicit representation for integer matrices 
that makes it possible to update as many as $\Omega(n^2)$ entries per 
operation in $o(n^2)$ worst-case time at the price of increasing the 
lookup time required to read a single entry.

\subsection{Dynamic Reevaluation of Polynomials over Boolean Matrices}
\label{se:tc-polynomials}

We now study the problem of maintaining the value of polynomials over 
Boolean matrices under updates of their variables.  We define these 
updates so that they can be useful later on for our original problem 
of dynamic Boolean matrix closure.  We first need some preliminary 
definitions.

\begin{definition}
\label{def:dynamic-objects}
Let $X$ be a data structure.  We denote by $X_i$ the value of $X$ at 
\define{Time} $i$, i.e., the value of $X$ after the $i$-th operation 
in a sequence of operations that modify $X$.  By convention, we assume 
that at time $0$ any numerical value in $X$ is zero.  In particular, 
if $X$ is a Boolean matrix, $X_0=0_n$.
\end{definition}

In the following definition we formally introduce our first problem on
dynamic matrices.

\begin{definition}
\label{def:polynomials}
Let ${\cal B}_{n}$ be the set of $n\times n$ Boolean matrices and let 
$$P=\sum_{a=1}^{h} T_{a}$$ be a polynomial\footnote{In the following, 
we omit specifying explicitly the dependence of a polynomial on its 
variables, and we denote by $P$ both the function 
$P(X_{1},\ldots,X_{k})$ and the value of this function for fixed 
values of $X_{1},\ldots,X_{k}$, assuming that the correct 
interpretation is clear from the context.} with $h$ terms defined over 
${\cal B}_{n}$, where each $$T_{a}=\prod_{b=1}^{k} X_{b}^{a}$$ has 
degree exactly $k$ and variables $X_{b}^{a}\in {\cal B}_n$ are 
distinct.  We consider the problem of maintaining a data structure 
{\tt P} for the polynomial $P$ under an intermixed sequence 
$\sigma=\langle {\tt P.Op}_1,\ldots,{\tt P.Op}_l\rangle$ of 
initialization, update, and query operations.  Each operation ${\tt 
P.Op}_j$ on the data structure {\tt P} can be either one of the 
following:

\begin{itemize}

\item {\tt P.Init}$(Z_{1}^{1},\ldots,Z_{k}^{h})$: perform the
initialization $X_{b}^{a}\leftarrow Z_{b}^{a}$ of the variables of
polynomial $P$, where each $Z_{b}^{a}$ is an $n\times n$ Boolean
matrix.

\item {\tt P.SetRow}$(i,\Delta X,X_{b}^{a})$: perform the row update
operation $X_{b}^{a}\leftarrow X_{b}^{a}+I_{\Delta X,i}$, where
$\Delta X$ is an $n\times n$ Boolean update matrix.  The operation
sets to $1$ the entries in the $i$-th row of variable $X_{b}^{a}$ of
polynomial $P$ as specified by matrix $\Delta X$.

\item {\tt P.SetCol}$(i,\Delta X,X_{b}^{a})$: perform the column
update operation $X_{b}^{a}\leftarrow X_{b}^{a}+J_{\Delta X,i}$, where
$\Delta X$ is an $n\times n$ Boolean update matrix.  The operation
sets to $1$ the entries in the $i$-th column of variable $X_{b}^{a}$
of polynomial $P$ as specified by matrix $\Delta X$.

\item {\tt P.LazySet}$(\Delta X,X_{b}^{a})$: perform the update
operation $X_{b}^{a}\leftarrow X_{b}^{a}+\Delta X$, where $\Delta X$
is an $n\times n$ Boolean update matrix.  The operation sets to $1$
the entries of variable $X_{b}^{a}$ of polynomial $P$ as
specified by matrix $\Delta X$.

\item {\tt P.Reset}$(\Delta X,X_{b}^{a})$: perform the update
operation $X_{b}^{a}\leftarrow X_{b}^{a}-\Delta X$, where $\Delta X$
is an $n\times n$ Boolean update matrix such that $\Delta X\subseteq
X_{b}^{a}$.  The operation resets to $0$ the entries of variable
$X_{b}^{a}$ of polynomial $P$ as specified by matrix $\Delta X$.

\item {\tt P.Lookup}$()$: answer a query about the value of $P$ by
returning an $n\times n$ Boolean matrix $Y_j$, such that $M_j\subseteq
Y_j\subseteq P_j$, where $M$ is an $n\times n$ Boolean matrix whose
value at time $j$ is defined as follows:
$$
M_j=\hspace{-3mm}\sum_{\scriptsize
\begin{array}{c}
1\le i\le j: \\
{\tt Op}_i\neq {\tt LazySet}\\
\end{array}}\hspace{-3mm} (P_{i}-P_{i-1})
$$
and $P_{i}$ is the value of polynomial $P$ at time $i$.  According to 
this definition, we allow the answer about the value of $P$ to be 
affected by one-side error.

\end{itemize}

\end{definition}

{\tt SetRow} and {\tt SetCol} are allowed to modify only the $i$-th 
row and the $i$-th column of variable $X_{b}^{a}$, respectively, while 
{\tt LazySet}, {\tt Reset} and {\tt Init} can modify any entries of 
$X_{b}^{a}$.  It is crucial to observe that in the operational 
definition of {\tt Lookup} we allow one-side errors in answering 
queries on the value of $P$.  In particular, in the answer there have 
to be no incorrect $1$'s and the error must be bounded: {\tt Lookup} 
has to return a matrix $Y$ that contains {\em at least} the $1$'s in 
$M$, and {\em no more} than the $1$'s in $P$.  As we will see later 
on, this operational definition simplifies the task of designing 
efficient implementations of the operations and is still powerful 
enough to be useful for our original problem of dynamic Boolean matrix 
closure.

The following lemma shows that the presence of errors is related to 
the presence of {\tt LazySet} operations in sequence $\sigma$.  In 
particular, it shows that, if no {\tt LazySet} operation is performed, 
then {\tt Lookup} makes no errors and returns the correct value of 
polynomial $P$.

\begin{lemma}
Let $P$ be a polynomial and let $\sigma=\langle {\tt
P.Op}_1,\ldots,{\tt P.Op}_k\rangle$ be a sequence of operations on
$P$.  If ${\tt Op}_i\neq {\tt LazySet}$ for all $1\le i\le j\le k$,
then $M_j=P_j$.
\end{lemma}

\begin{proof}
The proof easily follows by telescoping the sum that defines $M_j$: 
$M_j=P_{j}-P_{j-1}+P_{j-1}-P_{j-2}+\cdots+P_2-P_1+P_1-P_0=P_j-P_0=P_j.$
\end{proof}

Errors in the answers given by {\tt Lookup} may appear as soon as {\tt 
LazySet} operations are performed in sequence $\sigma$.  To explain 
how $M$ is defined mathematically, notice that $M_0=0_n$ by 
Definition~\ref{def:dynamic-objects} and $M$ sums up all the changes 
that the value of $P$ has undergone up to the $j$-th operation, except 
for the changes due to {\tt LazySet} operations, which are ignored.  
This means that, if any entry $P[x,y]$ flips from $0$ to $1$ or 
vice-versa due to an operation {\tt Op}$_{j}$ different from {\tt 
LazySet}, so does $M[x,y]$ and thus $Y[x,y]$.

As a side note, we remark that it is straightforward to extend the
results of this section to the general class of polynomials with terms
of different degrees and multiple occurrences of the same variable.

\gimmebreak

We now focus on the problem of implementing the operations introduced 
in Definition~\ref{def:polynomials}.  A simple-minded implementation 
of the operations on {\tt P} is the following:

\begin{itemize}

\item Maintain variables $X_{b}^a$, terms $T_{a}$, and a matrix $Y$ 
that contains the value of the polynomial.

\item Recompute from scratch $T_{a}$ and the value of 
$Y=P=T_{1}+\cdots+T_{h}$ after each {\tt Init}, {\tt SetRow}, {\tt 
SetCol} and {\tt Reset} that change $X_{b}^a$.

\item Do {\em nothing} after a {\tt LazySet} operation, except for 
updating $X_{b}^a$.  This means that $Y$ may be no longer equal to $P$ 
after the operation.

\item Let {\tt Lookup} return the maintained value of $Y$.

\end{itemize}

It is easy to verify that at any time $j$, i.e., after the $j$-th 
operation, {\tt Op}$_{j}\neq ${\tt LazySet} implies $Y=P$ and {\tt 
Op}$_j=${\tt LazySet} implies $Y=M$.  In other words, the value $Y$ 
returned by {\tt Lookup} oscillates between the exact value $P$ of the 
polynomial and the value $M$ obtained without considering {\tt 
LazySet} operations.

With the simple-minded implementation above, we can support {\tt Init} 
in $O(h\cdot k\cdot n^\omega+h\cdot n^2)$ time, {\tt SetRow} and {\tt 
SetCol} in $O(k\cdot n^\omega)$ time, {\tt Reset} in $O(k\cdot 
n^\omega+h\cdot n^2)$ time, and {\tt Lookup} and {\tt LazySet} in 
$O(n^2)$ time.

The remainder of this section provides more efficient solutions for 
the problem.  In particular, we present a data structure that supports 
{\tt Lookup} and {\tt LazySet} operations in $O(n^2)$ worst-case time, 
{\tt SetRow}, {\tt SetCol} and {\tt Reset} operations in $O(k\cdot 
n^2)$ amortized time, and {\tt Init} operations in $O(h\cdot k\cdot 
n^\omega+h\cdot n^2)$ worst-case time.  The space used is $O(h\cdot 
k^2\cdot n^2)$.  Before considering the general case where polynomials 
have arbitrary degree $k$, we focus on the special class of 
polynomials where $k=2$.

\subsubsection{Data Structure for Polynomials of Degree $k=2$}

We define a data structure for $P$ that allows us to maintain 
explicitly the value $Y_j$ of the matrix $Y$ at any time $j$ during a 
sequence $\langle {\tt P.Op}_1,\ldots,{\tt P.Op}_l\rangle$ of 
operations.  This makes it possible to perform {\tt Lookup} operations 
in optimal quadratic time.  We avoid recomputing from scratch the 
value of $Y$ after each update as in the simple-minded method, and we 
propose efficient techniques for propagating to $Y$ the effects of 
changes of variables $X_{b}^{a}$ due to {\tt SetRow}, {\tt SetCol} and 
{\tt Reset} operations.  In case of {\tt LazySet}, we only need to 
update the affected variables, leaving the other elements in the data 
structure unaffected.  This, of course, implies that after a {\tt 
LazySet} at time $j$, the maintained value $Y_{j}$ will be clearly not 
synchronized with the correct value $P_{j}$ of the polynomial.  Most 
technical difficulties of this section come just from this lazy 
maintenance of $Y_{j}$.

Our data structure for representing a polynomial of degree $2$ of the
form $P=X_1^1\cdot X_2^1+\ldots+X_1^h\cdot X_2^h$ is presented below.

\begin{datastructure}
\label{ds:poly-k=2}
We maintain the following elementary data structures with $O(h\cdot 
n^2)$ space:

\begin{enumerate}

\item $2h$ matrices $X_1^a$ and $X_2^a$ for $1\le a\le h$;

\item $h$ integer matrices $Prod_1,\ldots,Prod_h$ such that $Prod_{a}$ 
maintains a ``lazy'' count of the number of witnesses of the product 
$T_{a}=X_{1}^a\cdot X_{2}^a$.

\item an integer matrix $S$ such that $S[x,y]=|\{a:Prod_a[x,y]>0\}|$.
We assume that $Y_{j}[x,y]=1~\Leftrightarrow~S[x,y]>0$.

\item $2h$ integer matrices $LastFlip_X$, one for each matrix
$X=X_{b}^{a}$.  For any entry $X[x,y]=1$, $LastFlip_X[x,y]$ is the time
of the most recent operation that caused $X[x,y]$ to flip from $0$ to
$1$.  More formally: $$LastFlip_{X_j}[x,y]=\max_{1\le t\le
j}\{t~|~X_{t}[x,y]-X_{t-1}[x,y]=1\}$$
if $X_j[x,y]=1$, and is undefined otherwise;

\item $2h$ integer vectors $LastRow_X$, one for each matrix
$X=X_{b}^{a}$.  $LastRow_X[i]$ is the time of the last {\tt Init} or
{\tt SetRow} operation on the $i$-th row of $X$, and zero if no such
operation was ever performed.  More formally:
$$
LastRow_{X_j}[i]=\max_{1\le t\le j}\{0,~t~|~{\tt Op}_t={\tt Init}(\ldots)
~\vee~{\tt Op}_t={\tt SetRow}(i,\Delta X,X)\}
$$
We also maintain similar vectors $LastCol_X$;

\item a counter $Time$ of the number of performed operations;

\end{enumerate}
\end{datastructure}

Before getting into the full details of our implementation of
operations, we give an overview of the main ideas.  We consider how
the various operations should affect the data structure.  In
particular, we suppose that an operation changes any entries of
variable $X_{1}^a$ in a term $T_{a}=X_{1}^a\cdot X_{2}^a$, and we
define what our implementation should do on matrix $Prod_{a}$:

\begin{description}

\item [{\tt SetRow}/{\tt SetCol}:] if some entry $X_{1}^a[x,y]$ is 
flipping to $1$, then $y$ becomes a witness in the product 
$X_{1}^a\cdot X_{2}^a$ for any pair $x,z$ such that $X_{2}^a[y,z]=1$.  
Then we should put $y$ in the count $Prod_{a}[x,z]$, if it is not 
already counted.  Moreover, if some entry $X_{1}^a[x,y]$ was already 
$1$, but for some pair $x,z$ the index $y$ is not counted in 
$Prod_{a}[x,z]$, then we should put $y$ in the count $Prod_{a}[x,z]$.

\item [{\tt LazySet}:] if some entry $X_{1}^a[x,y]$ is flipping to 
$1$, then $y$ becomes a witness for any pair $x,z$ such that 
$X_{2}^a[y,z]=1$.  Then we should put $y$ in the count 
$Prod_{a}[x,z]$, if it is not already counted, {\em but we do not do 
this}.

\item [{\tt Reset}:] if some entry $X_{1}^a[x,y]$ is flipping to $0$, 
then $y$ is no longer a witness for all pairs $x,z$ such that 
$X_{2}^a[y,z]=1$.  Then we should remove $y$ from the count 
$Prod_{a}[x,z]$, if it is currently counted.

\end{description}

Note that after performing {\tt LazySet} there may be triples 
$(x,y,z)$ such that both $X_{1}^a[x,y]=1$ and $X_{2}^a[y,z]=1$, but 
$y$ is not counted in $Prod_{a}[x,z]$.  Now the problem is: is there 
any property that we can exploit to tell if a given $y$ is counted or 
not in $Prod_{a}[x,z]$ whenever both $X_{1}^a[x,y]=1$ and 
$X_{2}^a[y,z]=1$?

We introduce a predicate ${\cal P}_a(x,y,z)$, $1\leq x,y,z\leq n$,
such that ${\cal P}_a(x,y,z)$ is true if and only if the last time any
of the two entries $X_{1}^{a}[x,y]$ and $X_{2}^{a}[y,z]$ flipped from
$0$ to $1$ is before the time of the last update operation on the
$x$-th row or the $y$-th column of $X_{1}^{a}$ and the time of the
last update operation on the $y$-th row or the $z$-th column of
$X_{2}^{a}$.  In short:

\begin{center}
${\cal P}_a(x,y,z)~:=~$
$\max\{LastFlip_{X_{1}^{a}}[x,y],LastFlip_{X_{2}^{a}}[y,z]\}\le$
$\max\{LastRow_{X_{1}^{a}}[x], LastCol_{X_{1}^{a}}[y],
LastRow_{X_{2}^{a}}[y], LastCol_{X_{2}^{a}}[z]\}$
\end{center}

The property ${\cal P}_a$ answers our previous question and allows it 
to define the following invariant that we maintain in our data 
structure.  We remark that we do not need to maintain ${\cal P}_a$ 
explicitly in our data structure as it can be computed on demand in 
constant time by accessing $LastFlip$ and $LastRow$.

\begin{invariant}
\label{inv:polynomials}
For any term $T_a=X_{1}^{a}\cdot X_{2}^{a}$ in polynomial $P$, at any
time during a sequence of operations $\sigma$, the following invariant
holds for any pair of indices $x,z$:
\begin{center}
$Prod_a[x,z]=|\{y~:~X_{1}^{a}[x,y]=1~\wedge~X_{2}^{a}[y,z]=1~\wedge~{\cal
P}_a(x,y,z)\}|$
\end{center}
\end{invariant}

According to Invariant~\ref{inv:polynomials}, it is clear that the 
value of each entry $Prod_a[x,z]$ is a ``lazy'' count of the number of 
witnesses of the Boolean matrix product $T_a[x,z]=(X_{1}^{a}\cdot 
X_{2}^{a})[x,z]$.  Notice that, since 
$T_a[x,z]=1~\Leftrightarrow~\exists 
y~:~X_{1}^{a}[x,y]=1~\wedge~X_{2}^{a}[y,z]=1$, we have that 
$Prod_a[x,z]>0~\Rightarrow~T_a[x,z]=1$.  Thus, we may think of ${\cal 
P}_a$ as a ``relaxation'' property.

\medskip

We implement the operations introduced in
Definition~\ref{def:polynomials} as described next, assuming that the
operation $Time\leftarrow Time+1$ is performed just before each
operation:

\bigskip \noindent
\begin{minipage}{15cm}

\noindent{\large \tt Init}~\hrulefill~

\medskip

\begin{frameprog}{15cm}{\small}
\PROCEDURE\ {\tt Init}$(Z_{1}^{1},
Z_{2}^{1},\ldots,Z_{1}^{h}, Z_{2}^{h})$ \\
\N \BEGIN \\
\N \> \FOREACH\ $a$ \DO\ $X_{1}^{a}\leftarrow Z_{1}^{a}$;
$X_{2}^{a}\leftarrow Z_{2}^{a}$ \\
\N \> \{ initialize members $2$--$5$ of Data Structure~\ref{ds:poly-k=2} \} \\
\N \END \\
\end{frameprog}

\medskip

{\tt Init} assigns the value of variables $X_{1}^{a}$ and $X_{2}^{a}$ 
and initializes elements $2$--$5$ of Data Structure~\ref{ds:poly-k=2}.  
In particular, $LastFlip_X[x,y]$ is set to $Time$ for any $X[x,y]=1$ 
and the same is done for $LastRow[i]$ and $LastCol[i]$ for any $i$.  
$Prod_a$ is initialized by computing the product $X_{1}^{a}\cdot 
X_{2}^{a}$ in the ring of integers, i.e., looking at $X_{b}^{a}$ as 
integer matrices.

\end{minipage}

\bigskip \noindent
\begin{minipage}{15cm}

\noindent{\large \tt Lookup}~\hrulefill~

\medskip

\begin{frameprog}{15cm}{\small}
\FUNCTION\ {\tt Lookup}$()$ \\
\N \BEGIN \\
\N \> \RETURN\ $Y$ s.t. $Y[x,y]=1~\Leftrightarrow~S[x,y]>0$ \\
\N \END \\
\end{frameprog}

\medskip

{\tt Lookup} simply returns a binarized version $Y$ of matrix $S$
defined in Data Structure~\ref{ds:poly-k=2}.

\end{minipage}

\bigskip \noindent
\begin{minipage}{15cm}

\noindent{\large \tt SetRow}~\hrulefill~

\medskip

\begin{frameprog}{15cm}{\small}
\PROCEDURE\ {\tt SetRow}$(i,\Delta X,X_{b}^{a})$ \\
\N \BEGIN \\
\N \> $X_{b}^{a}\leftarrow X_{b}^{a}+I_{\Delta X,i}$ \\
\N \> \{update $LastFlip_{X_{b}^{a}}$\} \\
\N \> \IF\ $b=1$ \THEN \\
\N \> \> \FOREACH\ $x:X_{1}^{a}[i,x]=1$ \DO \\
\N \> \> \> \FOREACH\ $y:X_{2}^{a}[x,y]=1$ \DO \\
\N \> \> \> \> \IF\ \NOT\ ${\cal P}_a(i,x,y)$ \THEN \\
\N \> \> \> \> \> $Prod_a[i,y]\leftarrow Prod_a[i,y]+1$ \\
\N \> \> \> \> \> \IF\ $Prod_a[i,y]=1$ \THEN\ $S[i,y]\leftarrow S[i,y]+1$ \\
\N \> \ELSE \{$b=2$: similar to {\tt P.SetCol}$(i,\Delta X,X_{1}^{a})$\} \\
\N \> $LastRow_{X_{b}^{a}}[i]\leftarrow Time$ \\
\N \END \\
\end{frameprog}
\end{minipage}

\medskip

\noindent After performing an $i$-centered insertion in $X_{b}^{a}$ on 
line $2$ and after updating $LastFlip_{X_{b}^{a}}$ on line $3$, {\tt 
SetRow} checks on lines $5$--$7$ for any triple $(i,x,y)$ such that 
the property ${\cal P}_a(i,x,y)$ is still not satisfied, but will be 
satisfied thanks to line $11$, and increases $Prod_a$ and $S$ 
accordingly (lines $8$--$9$).

\bigskip \noindent
\begin{minipage}{15cm}

\noindent{\large \tt SetCol}~\hrulefill~

\medskip

\begin{frameprog}{15cm}{\small}
\PROCEDURE\ {\tt SetCol}$(i,\Delta X,X_{b}^{a})$ \\
\N \BEGIN \\
\N \> $X_{b}^{a}\leftarrow X_{b}^{a}+J_{\Delta X,i}$ \\
\N \> \{update $LastFlip_{X_{b}^{a}}$\} \\
\N \> \IF\ $b=1$ \THEN \\
\N \> \> \FOREACH\ $x:X_{1}^{a}[x,i]=1$ \DO \\
\N \> \> \> \FOREACH\ $y:X_{2}^{a}[i,y]=1$ \DO \\
\N \> \> \> \> \IF\ \NOT\ ${\cal P}_a(x,i,y)$ \THEN \\
\N \> \> \> \> \> $Prod_a[x,y]\leftarrow Prod_a[x,y]+1$ \\
\N \> \> \> \> \> \IF\ $Prod_a[x,y]=1$ \THEN\ $S[x,y]\leftarrow S[x,y]+1$ \\
\N \> \ELSE\ \{$b=2$: similar to {\tt P.SetRow}$(i,\Delta X,X_{1}^{a})$\} \\
\N \> $LastCol_{X_{1}^{a}}[i]\leftarrow Time$ \\
\N \END \\
\end{frameprog}
\end{minipage}

\medskip

\noindent Similar to {\tt SetRow}.

\bigskip \noindent
\begin{minipage}{15cm}

\noindent{\large \tt LazySet}~\hrulefill~

\medskip

\begin{frameprog}{15cm}{\small}
\PROCEDURE\ {\tt LazySet}$(\Delta X,X_{b}^{a})$ \\
\N \BEGIN \\
\N \> $X_{b}^{a}\leftarrow X_{b}^{a}+\Delta X$ \\
\N \> \{update $LastFlip_{X_{b}^{a}}$\} \\
\N \END \\
\end{frameprog}
\end{minipage}

\medskip

\noindent {\tt LazySet} simply sets to $1$ any entries in $X_{b}^{a}$ and
updates $LastFlip_{X_{b}^{a}}$.  We remark that no other object in the
data structure is changed.

\bigskip \noindent
\begin{minipage}{15cm}

\noindent{\large \tt Reset}~\hrulefill~

\medskip

\begin{frameprog}{15cm}{\small}
\PROCEDURE\ {\tt Reset}$(\Delta X,X_{b}^{a})$ \\
\N \BEGIN \\
\N \> \IF\ $b=1$ \THEN \\
\N \> \> \FOREACH\ $x,y:\Delta X[x,y]=1$ \DO \\
\N \> \> \> \IF\ $\max\{LastRow_{X_{1}^{a}}[x],
                        LastCol_{X_{1}^{a}}[y]\}\ge LastFlip_{X_{1}^{a}}[x,y]$ \THEN\ \\
\N \> \> \> \> \FOREACH\ $z:X_{1}^{a}[y,z]=1$ \DO \\
\N \> \> \> \> \> \IF\ ${\cal P}_a(x,y,z)$ \THEN \\
\N \> \> \> \> \> \> $Prod_a[x,z]\leftarrow Prod_a[x,z]-1$ \\
\N \> \> \> \> \> \> \IF\ $Prod_a[x,z]=0$ \THEN\ $S[x,z]\leftarrow S[x,z]-1$ \\
\N \> \> \> \ELSE\ \{~here~$\max\{LastRow_{X_{1}^{a}}[x],
                        LastCol_{X_{1}^{a}}[y]\}<
                        LastFlip_{X_{1}^{a}}[x,y]$~\} \\
\N \> \> \> \> \FOREACH\ $z:X_{1}^{a}[y,z]=1~\wedge~LastCol_{X_{2}^{a}}[z]>LastFlip_{X_{1}^{a}}[x,y]$ \DO \\
\N \> \> \> \> \> \IF\ ${\cal P}_a(x,y,z)$ \THEN \\
\N \> \> \> \> \> \> $Prod_a[x,z]\leftarrow Prod_a[x,z]-1$ \\
\N \> \> \> \> \> \> \IF\ $Prod_a[x,z]=0$ \THEN\ $S[x,z]\leftarrow S[x,z]-1$ \\

\N \> \ELSE\ \{$b=2$ similar to $b=1$\} \\
\N \> $X_{b}^{a}\leftarrow X_{b}^{a}-\Delta X$ \\
\N \END \\
\end{frameprog}
\end{minipage}

\medskip

\noindent In lines $2$-$14$, using $LastRow_{X_{b}^{a}}$,
$LastCol_{X_{b}^{a}}$, and $LastFlip_{X_{b}^{a}}$, {\tt Reset} updates
$Prod_a$ and $S$ so as to maintain Invariant~\ref{inv:polynomials}.
Namely, for each reset entry $(x,y)$ specified by $\Delta X$ (line
$3$), it looks for triples $(x,y,z)$ such that ${\cal P}(x,y,z)$ is
going to be no more satisfied due to the reset of $X_{b}^{a}[x,y]$ to
be performed (lines $5$--$6$ and lines $10$--$11$); $Prod_a$ and $S$
are adjusted accordingly (lines $7$--$8$ and lines $12$-$13$).

\noindent The distinction between the two cases 
$\max\{LastRow_{X_{1}^{a}}[x], LastCol_{X_{1}^{a}}[y]\}\ge 
LastFlip_{X_{1}^{a}[x,y]}$ and $\max\{LastRow_{X_{1}^{a}}[x], 
LastCol_{X_{1}^{a}}[y]\}< LastFlip_{X_{1}^{a}[x,y]}$ in line $4$ and 
in line $9$, respectively, is important to achieve fast running times 
as it will be discussed in the proof of 
Theorem~\ref{th:polynomials-complexity}.  Here we only point out that 
if the test in line $4$ succeeds, then we can scan any $z$ s.t.  
$X_{1}^{a}[y,z]=1$ without affecting the running time.  If this is 
not the case, then we need to process {\em only} indices $z$ such that 
the test $LastCol_{X_{2}^{a}}[y,z]>LastFlip_{X_{1}^{a}}[x,y]$ is 
satisfied, and avoid scanning other indices.  For this reason line 
$10$ must be implemented very carefully by maintaining indices $z$ in 
a list and by using a move-to-front strategy that brings index $z$ to 
the front of the list as any operation {\tt Init}$(\ldots)$, {\tt 
SetRow}$(z,\ldots)$ or {\tt SetCol}$(z,\ldots)$ is performed on $z$.  
In this way indices are sorted according to the dates of operations on 
them.

\noindent As last step, {\tt Reset} resets the entries of $X_{b}^{a}$
as specified by $\Delta X$ (line $15$).

\gimmebreak

The correctness of our implementation of operations {\tt Init}, {\tt
SetRow}, {\tt SetCol}, {\tt LazySet}, {\tt Reset} and {\tt Lookup} is
discussed in the following theorem.

\begin{theorem}
\label{th:polynomials-correctness}
At any time $j$, {\tt Lookup} returns a matrix $Y_j$ that satisfies
the relation $M_{j}\subseteq Y_j\subseteq P_{j}$ as in
Definition~\ref{def:polynomials}.
\end{theorem}

\begin{proof}
We first remind that $Y$ is the binarized version of $S$ as follows
from the implementation of {\tt Lookup}.

To prove that $Y\subseteq P$, observe that {\tt SetRow} increases
$Prod_a[i,y]$ (line $8$), and possibly $S$ (line $9$), only if both
$X_{1}^{a}[i,x]=1$ and $X_{2}^{a}[x,y]=1$: this implies that
$T_a[i,y]=1$ and $P[i,y]=1$.

To prove that $M\subseteq Y$, notice that at time $j$ {\em after} 
performing an operation {\tt Op$_{j}$=SetRow}$(i,\Delta X,X_{b}^{a})$ 
on the $i$-th row of $X_{1}^{a}$, ${\cal P}(i,x,y)$ is satisfied for 
any triple $(i,x,y)$ such that $X_{1}^{a}[i,x]=1$ and 
$X_{2}^{a}[x,y]=1$ thanks to the operation 
$LastRow_{X_{b}^{a}}[i]\leftarrow Time$ (line $11$).  For $X_{2}^{a}$ 
the proof is analogous.  Now, all such triples $(i,x,y)$ are 
enumerated by {\tt SetRow} (lines $5$--$6$): for each of them such 
that ${\cal P}_{a}(i,x,y)$ was false at time $j-1$, $Prod_a[i,y]$ is 
increased and possibly $S[i,y]$ is increased as well (lines $7$-$9$).  
If $P[i,y]$ flips from $0$ to $1$, then necessarily $X_{1}^{a}[i,x]$ 
flips from $0$ to $1$ for some $x$, and then, as stated above w.r.t.  
${\cal P}_{a}$, $Prod_a[i,y]$ gets increased.  Thus, recalling that 
$Y$ is the binarized version of $S$, we have for any $y$:
$$P_{j}[i,y]-P_{j-1}[i,y]=1~\Rightarrow~Y_j[i,y]-Y_{j-1}[i,y]=1.$$
From the definition of $M$ in Definition~\ref{def:polynomials} we have
that:
$$M_{j}[i,y]-M_{j-1}[i,y]=1~\Leftrightarrow~P_{j}[i,y]-P_{j-1}[i,y]=1.$$
This proves the relation $M\subseteq Y$.  A similar argument is valid 
also for {\tt SetCol}, while {\tt LazySet} does not affect $S$ at all.

To complete the proof we remark that $Y=P$ just after any {\tt Init} 
operation and that {\tt Reset} leaves the data structure as if reset 
entries were never set to $1$.  Indeed, {\tt Reset} can be viewed as a 
sort of ``undo'' procedure that cancels the effects of previous {\tt 
SetRow}, {\tt SetCol} or {\tt Init} operations.
\end{proof}

We now analyze the complexity of our implementation of the operations
on polynomials.

\begin{theorem}
\label{th:polynomials-complexity}
Any {\tt Lookup}, {\tt SetRow}, {\tt SetCol} and {\tt LazySet}
operation requires $O(n^2)$ time in the worst case.  Any {\tt Init}
requires $O(h\cdot n^\omega+h\cdot n^2)$ worst-case time, where
$\omega$ is the exponent of matrix multiplication.  The cost of any
{\tt Reset} operation can be charged to previous {\tt SetRow}, {\tt
SetCol} and {\tt Init} operations.  The maximum cost charged to each
{\tt Init} is $O(h\cdot n^3)$.  The space required is $O(h\cdot n^2)$.
\end{theorem}

\begin{proof}
It is straightforward to see from the pseudocode of the operations
that any {\tt SetRow}, {\tt SetCol} and {\tt LazySet} operation
requires $O(n^2)$ time in the worst case.

{\tt Init} takes $O(h\cdot n^\omega+h\cdot n^2)$ in the worst case: in 
more detail, each $Prod_a$ can be directly computed via matrix 
multiplication and any other initialization step requires no more than 
$O(n^2)$ worst-case time.

To prove that the cost of any {\tt Reset} operation can be charged to
previous {\tt SetRow}, {\tt SetCol} and {\tt Init} operations, we use
a potential function $$\Phi_a=\sum_{x,y}Prod_a[x,y]$$ associated to
each term $T_a$ of the polynomial.  From the relation:
$$Prod_a[x,z]=|\{y~:~X_{1}^{a}[x,y]=1~\wedge~X_{2}^{a}[y,z]=1~\wedge~{\cal
P}_a(x,y,z)\}|$$ given in Invariant~\ref{inv:polynomials}, it follows
that $0\leq Prod_a[x,z]\leq n$ for all $x,z$.  Thus, $0\leq\Phi_a\leq
n^3$.

Now, observe that {\tt SetRow} increases $\Phi_a$ by at most $n^2$ 
units per operation, while {\tt Init} increases $\Phi_a$ by at most 
$n^3$ units per operation.  Note that $LazySet$ does not affect 
$\Phi_a$.  We can finally address the case of {\tt Reset} operations.  
Consider the distinction between the two cases 
$\max\{LastRow_{X_{1}^{a}}[x], LastCol_{X_{1}^{a}}[y]\}$ $\ge 
LastFlip_{X_{1}^{a}[x,y]}$ in line $4$ and 
$\max\{LastRow_{X_{1}^{a}}[x],LastCol_{X_{1}^{a}}[y]\}< 
LastFlip_{X_{1}^{a}[x,y]}$ in line $9$.  In the first case, we can 
charge the cost of processing any triple $(x,y,z)$ to some previous 
operation on the $x$-th row of $X_{1}^{a}$ or to some previous 
operation on the $y$-th column of $X_{1}^{a}$; in the second case, we 
consider only those $(x,y,z)$ for which some operation on the $z$-th 
column of $X_{2}^{a}[y,z]$ was performed {\em after} both 
$X_{1}^{a}[x,y]$ and $X_{2}^{a}[y,z]$ were set to $1$.  In both cases, 
any {\tt Reset} operation decreases $\Phi_a$ by at most $n$ units for 
each reset entry of $X_{b}^{a}$, and this can be charged to previous 
operations which increased $\Phi_a$.
\end{proof}

The complex statement of the charging mechanism encompasses the
dynamics of our data structure.  In particular, we allow {\tt Reset}
operations to charge up to a $O(n^3)$ cost to a single {\tt Init}
operation.  Thus, in an arbitrary mixed sequence with any number of
{\tt Init}, {\tt Reset} takes $O(n^3)$ amortized time per update. If,
however, we allow {\tt Init} operations to appear in $\sigma$ only
every $\Omega(n)$ {\tt Reset} operations, the bound for {\tt Reset}
drops down to $O(n^2)$ amortized time per operation.

As a consequence of Theorem~\ref{th:polynomials-complexity}, we have 
the following corollaries that refine the analysis of the running time 
of {\tt Reset} operations.

\begin{corollary}
If we perform just one {\tt Init} operation in a sequence $\sigma$ of
length $\Omega(n)$, or more generally one {\tt Init} operation every
$\Omega(n)$ {\tt Reset} operations, then the amortized cost of {\tt
Reset} is $O(n^2)$ per operation.
\end{corollary}

\begin{corollary}
If we perform just one {\tt Init} operation in a sequence $\sigma$ of 
length $\Omega(n^2)$, or more generally one {\tt Init} operation every 
$\Omega(n^2)$ {\tt Reset} operations, and no operations {\tt SetRow} 
and {\tt SetCol}, then the amortized cost of {\tt Reset} is $O(n)$ per 
operation.
\end{corollary}

In the following, we show how to extend the previous techniques in 
order to deal with the general case of polynomials of degree $k>2$.

\newpage

\subsubsection{Data Structure for Polynomials of Degree $k>2$}

To support terms of degree $k>2$ in $P$, we consider an equivalent 
representation $\widehat{P}$ of $P$ such that the degree of each term 
is $2$.  This allows us to maintain a data structure for $\widehat{P}$ 
with the operations defined in the previous paragraph.

\begin{lemma}
\label{le:equiv-poly}
Consider a polynomial 
$$P=\sum_{a=1}^{h}T_a=\sum_{a=1}^{h}X_{1}^{a}\cdots
X_{k}^{a}$$
with $h$ terms where each term $T_a$ has degree exactly $k$ and
variables $X_{b}^{a}$ are Boolean matrices.  Let $\widehat{P}$ be the
polynomial over Boolean matrices of degree $2$ defined as
$$
\widehat{P}=\sum_{a=1}^{h}\sum_{b=0}^{k}L_{b,b-1}^{a}\cdot R_{b,k-b-1}^{a}
$$
where $L_{b,j}^{a}$ and $R_{b,j}^{a}$ are polynomials over Boolean
matrices of degree $\le 2$ defined as

\bigskip

\noindent \hspace{25mm}$
L_{b,j}^{a} =
\left\{
\begin{array}{ll}
    X_{b-j}^{a}\cdot L_{b,j-1}^{a} & if\ j \in[0,b-1] \\
    I_{n} & if\ j=-1 \\
\end{array}
\right.
$

\medskip

\noindent \hspace{25mm}$
R_{b,j}^{a} =
\left\{
\begin{array}{ll}
    R_{b,j-1}^{a}\cdot X_{b+1+j}^{a} & if\ j\in[0,k-b-1] \\
    I_{n} & if\ j=-1 \\
\end{array}
\right.
$

\bigskip

\noindent Then $P={\widehat P}$.
\end{lemma}

\begin{proof}
To prove the claim, it suffices to check that
$$T_a=\sum_{b=0}^{k}L_{b,b-1}^{a}\cdot R_{b,k-b-1}^{a}$$
Unrolling the recursion for $L_{b,b-1}^{a}$, we obtain:
$$L_{b,b-1}^{a}=X_{1}^{a}\cdot L_{b,b-2}^{a}=X_{1}^{a}\cdot
X_{2}^{a}\cdot L_{b,b-3}^{a}=\cdots=X_{1}^{a}\cdot X_{2}^{a}\cdots
X_{b}^{a}\cdot I_n$$ Likewise, $R_{b,k-b-1}^{a}=I_n\cdot
X_{b+1}^{a}\cdots X_{k}^{a}$ holds.  Thus, by idempotence of the
closed semiring of Boolean matrices,
we finally have:
$$\sum_{b=0}^{k}L_{b,b-1}^{a}\cdot
R_{b,k-b-1}^{a}=\sum_{b=0}^{k}X_{1}^{a}\cdots X_{b}^{a}\cdot
X_{b+1}^{a}\cdots X_{k}^{a}=X_{1}^{a}\cdots
X_{k}^{a}=T_a.$$
\end{proof}

Since $\widehat{P}$, $L_{b,j}^{a}$ and $R_{b,j}^{a}$ are all 
polynomials of degree $\le 2$, they can be represented and maintained 
efficiently by means of instances of Data Structure~\ref{ds:poly-k=2}.  
Our data structure for maintaining polynomials of degree $>2$ is 
presented below:

\begin{datastructure}
\label{ds:poly-hi-degr}
We maintain explicitly the $k^2$ polynomials $L_{b,j}^{a}$ and 
$R_{b,j}^{a}$ with instances of Data Structure~\ref{ds:poly-k=2}.  We 
also maintain polynomial $\widehat{P}$ with an instance $Y$ of Data 
Structure~\ref{ds:poly-k=2}.
\end{datastructure}

We now consider how to support {\tt SetRow}, {\tt SetCol}, {\tt
LazySet}, {\tt Reset}, {\tt Init} and {\tt Lookup} in the case of
arbitrary degree.  We denote by ${\tt SetRow}_{k=2}$ and ${\tt
SetCol}_{k=2}$ the versions of {\tt SetRow} and {\tt SetCol}
implemented for $k=2$.

\begin{figure}
\centerline{ { \epsfxsize=15cm \epsffile{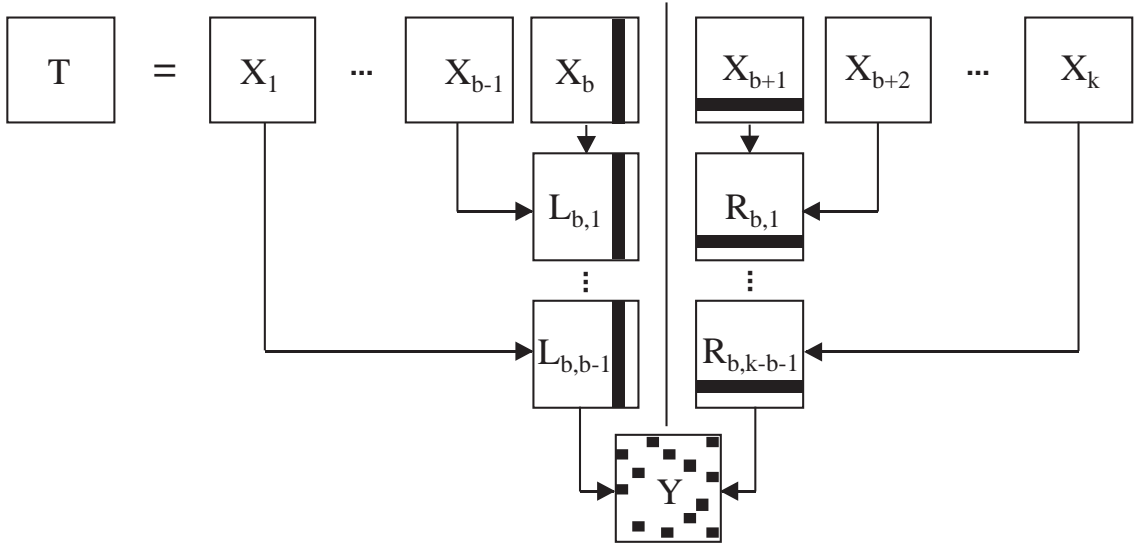}} }
\caption{Revealing new $1$'s in $Y$ while updating a term $T$ of 
degree $k$ by means of a {\tt SetCol} operation on variable $X_{b}$ or 
by means of a {\tt SetRow} operation on variable $X_{b+1}$.}
\label{fi:tc-hdegr-poly}
\end{figure}

\bigskip \noindent
\begin{minipage}{15cm}

\noindent{\large \tt SetCol, SetRow}~\hrulefill~

\medskip

\begin{frameprog}{15cm}{\small}
\PROCEDURE\ {\tt SetCol}$(i,\Delta X,X_{b}^{a})$ \\
\N \BEGIN \\
\N \> $X_{b}^{a}\leftarrow X_{b}^{a}+J_{\Delta X,i}$ \\
\N \> \FOR\ $j\leftarrow 1$ \TO\ $b-1$ \DO \\
\N \> \> ${\tt L_{b,j}^{a}.SetCol}_{k=2}(i,\Delta L_{b,j-1}^{a},L_{b,j-1}^{a})$ \{ it
holds $L_{b,j}^{a}=X_{b-j}^{a}\cdot L_{b,j-1}^{a}$ \}\\
\N \> \FOR\ $j\leftarrow 1$ \TO\ $k-b-1$ \DO \\
\N \> \> ${\tt R_{b,j}^{a}.SetRow}_{k=2}(i,\Delta R_{b,j-1}^{a},R_{b,j-1}^{a})$ \{ it
holds $R_{b,j}^{a}=R_{b,j-1}^{a}\cdot X_{b+1+j}^{a}$ \}\\
\N \> ${\tt Y}.{\tt SetCol}_{k=2}(i,\Delta L_{b,b-1}^{a},L_{b,b-1}^{a})$ \\
\N \> ${\tt Y}.{\tt SetRow}_{k=2}(i,\Delta R_{b,k-b-1}^{a},R_{b,k-b-1}^{a})$ \\
\N \> \FOR\ $j\leftarrow 1$ \TO\ $k-b$ \DO\ ${\tt L_{b+j,j}^a.LazySet}(\Delta X_{b}^a,X_{b}^a)$ \\
\N \> \FOR\ $j\leftarrow 1$ \TO\ $b-2$ \DO\ ${\tt R_{b-j-1,j}^a.LazySet}(\Delta X_{b}^a,X_{b}^a)$ \\
\N \END \\
\end{frameprog}
\end{minipage}

\medskip

\noindent The main idea behind {\tt SetCol} is to exploit associativity of 
Boolean matrix multiplication in order to propagate changes of 
intermediate polynomials that are always limited to a row or a column 
and thus can be efficiently handled by means of operations like {\tt 
SetRow}$_{k=2}$ and {\tt SetCol}$_{k=2}$.

\smallskip

In lines $3$--$4$ {\tt SetCol} propagates via ${\tt SetCol}_{k=2}$ the 
changes of the $i$-th column of $X_{b}^{a}$ to $L_{b,1}^{a}$, then the 
changes of the $i$-th column of $L_{b,1}^{a}$ to $L_{b,2}^{a}$, and so 
on through the recursive decomposition:

\medskip

\begin{minipage}{15cm}
$$
\begin{array}{llllr}
L_{b,0}^{a}&=&X_{b}^{a}  \cdot I_n        &=&X_{b}^{a} \\
L_{b,1}^{a}&=&X_{b-1}^{a}\cdot L_{b,0}^{a}&=&X_{b-1}^{a}\cdot X_{b}^{a} \\
L_{b,2}^{a}&=&X_{b-2}^{a}\cdot L_{b,1}^{a}&=&X_{b-2}^{a}\cdot X_{b-1}^{a}\cdot X_{b}^{a} \\
~~~\vdots & & ~~~~~~\vdots & & \vdots~~~~~~~~~ \\
L_{b,b-1}^{a}&=&X_{1}^{a}\cdot L_{b,b-2}^{a}&=& X_{1}^{a}\cdots X_{b-2}^{a}\cdot X_{b-1}^{a}\cdot X_{b}^{a}\\
\end{array}
$$
\end{minipage}

\medskip

Likewise, in lines $5$--$6$ it propagates via ${\tt SetRow}_{k=2}$ a
null matrix of changes of the $i$-th row of $X_{b+1}^{a}$ to
$R_{b,1}^{a}$, then the changes (possibly none) of the $i$-th row of
$R_{b,1}^{a}$ (due to the late effects of some previous {\tt LazySet})
to $R_{b,2}^{a}$, and so on through the recursive decomposition:

\medskip

\begin{minipage}{15cm}
$$
\begin{array}{lllll}
R_{b,0}^{a}&=&I_n\cdot X_{b}^{a}          &=&X_{b+1}^{a} \\
R_{b,1}^{a}&=&R_{b,0}^{a}\cdot X_{b+1}^{a}&=&X_{b+1}^{a}\cdot X_{b+2}^{a} \\
R_{b,2}^{a}&=&R_{b,1}^{a}\cdot X_{b+2}^{a}&=&X_{b+1}^{a}\cdot X_{b+2}^{a}\cdot X_{b+3}^{a} \\
~~~\vdots & & ~~~~~~\vdots & & ~~~~~~~~~\vdots \\
R_{b,k-b-1}^{a}&=&R_{b,k-b-2}^{a}\cdot X_{k}^{a}&=& X_{b+1}^{a}\cdot X_{b+2}^{a}\cdot X_{b+3}^{a}\cdots X_{k}^{a}\\
\end{array}
$$
\end{minipage}

\medskip

We remark that both loops in lines $3$--$4$ and in lines $5$--$6$ 
reveal, gather and propagate any $1$'s that appear in the intermediate 
polynomials due to the late effects of some previous {\tt LazySet}.  
In particular, even if the presence of lines $5$--$6$ may seem strange 
because $\Delta X_{b+1}^{a}=0_n$, these lines are executed just for 
this reason.

Finally, in lines $7$--$8$ changes of $L_{b,b-1}^{a}$ and 
$R_{b,k-b-1}^{a}$ are propagated to $Y$, which represents the 
maintained value of $\widehat{P}$, and in lines $9$--$10$ new $1$'s 
are lazily inserted in any other polynomials that feature $X_{b}^{a}$ 
as a variable.

\medskip

We omit the pseudocode for {\tt SetRow} because it is similar to {\tt 
SetCol}.

\bigskip \noindent
\begin{minipage}{15cm}

\noindent{\large \tt Reset, LazySet, Init, Lookup}~\hrulefill~

\medskip

{\tt Reset}$(\Delta X,X_{b}^{a})$ can be supported by propagating via 
{\tt Reset}$_{k=2}$ any changes of $X_{b}^{a}$ to any intermediate 
polynomial $L_{u,v}^{a}$ and $R_{u,v}^{a}$ that contains it, then 
changes of such polynomials to any polynomials which depend on them 
and so on up to $Y$.

\medskip

{\tt LazySet}$(\Delta X,X_{b}^{a})$ can be supported by performing 
{\tt LazySet}$_{k=2}$ operations on each polynomial $L_{u,v}^{a}$ and 
$R_{u,v}^{a}$ that contains $X_{b}^{a}$.

\medskip

{\tt Init}$(Z_{1}^{1},\ldots,Z_{k}^{h})$ can be supported by invoking 
{\tt Init}$_{k=2}$ on each polynomial $L_{u,v}^{w}$, $R_{u,v}^{w}$ and 
by propagating the intermediate results up to $Y$.

\medskip

{\tt Lookup}$()$ can be realized by returning the maintained value 
$Y$ of $\widehat{P}$.

\end{minipage}

\bigskip

To conclude this section, we discuss the correctness and the 
complexity of our operations in the case of polynomials of arbitrary 
degree.

\begin{theorem}
At any time $j$, {\tt Lookup} returns a matrix $Y_j$ that satisfies 
the relation $M_{j}\subseteq Y_j\subseteq P_{j}$ as in 
Definition~\ref{def:polynomials}.
\end{theorem}

\begin{proof}
Since $\widehat{P}=P$ by Lemma~\ref{le:equiv-poly}, we prove that: 
$$\widehat{P}_{j}\supseteq Y_{j}\supseteq 
M_{j}=\hspace{-3mm}\sum_{\scriptsize
\begin{array}{c}
1\le i\le j: \\
{\tt Op}_i\neq {\tt LazySet}\\
\end{array}}\hspace{-3mm} (\widehat{P}_{i}-\widehat{P}_{i-1}).
$$

To this aim, it is sufficient to prove that any $1$ that appears (or 
disappears) in the correct value of $\widehat{P}$ due to an operation 
different from {\tt LazySet} appears (or disappears) in $Y$ as well, 
and that any entry of $Y$ equal to $1$ is also equal to $1$ in 
$\widehat{P}$.

\begin{itemize}
\item {\tt SetCol}/{\tt SetRow}: assume a {\tt SetCol} operation is 
performed on the $i$-th column of variable $X_{b}^a$ (see 
Figure~\ref{fi:tc-hdegr-poly}).  By induction, we assume that all new 
$1$'s are correctly revealed in the $i$-th column of our data 
structure for $L_{b,j}^a$ after the $j$-th iteration of {\tt 
SetCol}$_{k=2}$ in line $4$.  Notice that $\Delta L_{b,j}^a=J_{\Delta 
L_{b,j}^a,i}$, that is changes of $L_{b,j}^a$ are limited to the 
$i$-th column: this implies that these changes can be correctly 
propagated by means of a {\tt SetCol} operation to any polynomial that 
features $L_{b,j}^a$ as a variable.  As a consequence, by 
Theorem~\ref{th:polynomials-correctness}, the $j+1$-th iteration of 
{\tt SetCol}$_{k=2}$ in line $4$ correctly reveals all new $1$'s in 
our data structure for $L_{b,j+1}^a$, and again these new $1$'s all 
lie on its $i$-th column.  Thus, at the end of the loop in lines 
$3$--$4$, all new $1$'s appear correctly in the $i$-th column of 
$L_{b,b-1}^a$.  Similar considerations apply also for $R_{b,k-b-1}^a$.  
To prove that lines $7$--$8$ insert correctly in $Y$ all new $1$'s 
that appear in $\widehat{P}$ and that $Y\subseteq \widehat{P}$ we use 
again Theorem~\ref{th:polynomials-correctness} and the fact that any 
$1$ that appears in $\widehat{P}$ also appears in $L_{b,b-1}^a\cdot 
R_{b,k-b-1}^a$.  Indeed, for any entry $\widehat{P}[x,y]$ that flips 
from $0$ to $1$ due to a change of the $i$-th column of $X_{b}^a$ or 
the $i$-th row of $X_{b+1}^a$ there is a sequence of indices 
$x=u_{0},u_{1},\ldots,u_{b-1},u_{b}=i,u_{b+1},\ldots,u_{k-1},u_{k}=y$ 
such that $X_{j}^a[u_{j-1},u_{j}]=1$, $1\le j\le k$, and either one of 
$X_{b}^a[u_{b-1},i]$ or $X_{b+1}^a[i,u_{b+1}]$ just flipped from $0$ 
to $1$ due to the {\tt SetRow}/{\tt SetCol} operation.  The proof for 
{\tt SetRow} is completely analogous.

\item {\tt Reset}: assume a {\tt Reset} operation is performed on 
variable $X_{b}^a$.  As {\tt Reset}$_{k=2}$ can reset any subset of 
entries of variables, and not only those lying on a row or a column as 
in the case of {\tt SetRow}$_{k=2}$ and {\tt SetCol}$_{k=2}$, the 
correctness of propagating any changes of $X_{b}^a$ to the polynomials 
that depend on it easily follows from 
Theorem~\ref{th:polynomials-correctness}.

\item {\tt Init}: each {\tt Init} operation recomputes from scratch 
all polynomials in Data Structure~\ref{ds:poly-hi-degr}.  Thus 
$Y=\widehat{P}$ after each {\tt Init} operation.

\end{itemize}

\end{proof}

\begin{theorem}
\label{th:polynomials2-complexity}
Any {\tt Lookup} and {\tt LazySet} operation requires $O(n^2)$ time in 
the worst case.  Any {\tt SetRow} and {\tt SetCol} operation requires 
$O(k\cdot n^2)$ amortized time, and any {\tt Init} operation takes 
$O(h\cdot k\cdot n^\omega+h\cdot n^2)$ worst-case time.  The cost of 
any {\tt Reset} operation can be charged to previous {\tt SetRow}, 
{\tt SetCol} and {\tt Init} operations.  The maximum cost charged to 
each {\tt Init} is $O(h\cdot k\cdot n^3)$.  The space required is 
$O(h\cdot k^2\cdot n^2)$.
\end{theorem}

\begin{proof}
The proof easily follows from Theorem~\ref{th:polynomials-complexity}.
\end{proof}

\smallskip
\noindent As in the previous paragraph, we have the following corollaries.

\begin{corollary}
If we perform just one {\tt Init} operation in a sequence $\sigma$ of
length $\Omega(n)$, or more generally one {\tt Init} operation every
$\Omega(n)$ {\tt Reset} operations, then the amortized cost of {\tt
Reset} is $O(k\cdot n^2)$ per operation.
\end{corollary}

\begin{corollary}
If we perform just one {\tt Init} operation in a sequence $\sigma$ of
length $\Omega(n^2)$, or more generally one {\tt Init} operation every
$\Omega(n^2)$ {\tt Reset} operations, and we perform no operations
{\tt SetRow} and {\tt SetCol}, then the amortized cost of {\tt Reset}
is $O(k\cdot n)$ per operation.
\end{corollary}

\subsection{Maintaining Dynamic Matrices over Integers}
\label{se:tc-subquadraticmatrices}

In this section we study the problem of finding an implicit 
representation for a matrix of integers that makes it possible to 
support simultaneous updates of multiple entries of the matrix very 
efficiently at the price of increasing the lookup time required to 
read a single entry.  This problem on dynamic matrices will be central 
to designing the first subquadratic algorithm for fully dynamic 
transitive closure that will be described in 
Section~\ref{se:tc-subquad}.  We formally define the problem as 
follows:

\begin{definition}
\label{def:subquad-matrix}
Let $M$ be an $n\times n$ integer matrix.  We consider the problem of 
performing an intermixed sequence $\sigma=\langle {\tt 
M.Op}_1,\ldots,{\tt M.Op}_l\rangle$ of operations on $M$, where each 
operation ${\tt M.Op}_j$ can be either one of the following:

\begin{itemize}

\item {\tt M.Init}$(X)$: perform the initialization $M\leftarrow X$,
where $X$ is an $n\times n$ integer matrix.

\item {\tt M.Update}$(J,I)$: perform the update operation $M\leftarrow
M+J\cdot I$, where $J$ is an $n\times 1$ column integer vector, and
$I$ is a $1\times n$ row integer vector.  The product $J\cdot I$ is
an $n\times n$ matrix defined for any $1\le x,y\le n$ as: $$(J\cdot
I)[x,y]=J[x]\cdot I[y]$$

\item {\tt M.Lookup}$(x,y)$: return the integer value $M[x,y]$.

\end{itemize}

\end{definition}

It is straightforward to observe that {\tt Lookup} can be supported in
unit time and operations {\tt Init} and {\tt Update} in $O(n^2)$
worst-case time by explicitly performing the algebraic operations
specified in the previous definition.

In the following we show that, if one is willing to give up unit time
for {\tt Lookup} operations, it is possible to support {\tt Update} in
$O(n^{\omega(1,\epsilon,1)-\epsilon})$ worst-case time for each update
operation, for any $\epsilon$, $0\leq\epsilon\leq 1$, where
$\omega(1,\epsilon,1)$ is the exponent of the multiplication of an
$n\times n^{\epsilon}$ matrix by an $n^{\epsilon}\times n$ matrix.
Queries on individual entries of $M$ are answered in $O(n^\epsilon)$
worst-case time via {\tt Lookup} operations and {\tt Init} still takes
$O(n^2)$ worst-case time.

We now sketch the main ideas behind the algorithm.  We follow a simple 
lazy approach: we log at most $n^\epsilon$ update operations without 
explicitly computing them and we perform a global reconstruction of 
the matrix every $n^\epsilon$ updates.  The reconstruction is done 
through fast rectangular matrix multiplication.  This yields an 
implicit representation for $M$ which requires us to run through 
logged updates in order to answer queries about entries of $M$.

\subsubsection*{Data Structure}

We maintain the following elementary data structures with $O(n^2)$ 
space:

\begin{itemize}

\item an $n\times n$ integer matrix $Lazy$ which maintains a lazy
representation of $M$;

\item an $n\times n^\epsilon$ integer matrix $Buf_J$ in which we
buffer update column vectors $J$;

\item an $n^\epsilon\times n$ integer matrix $Buf_I$ in which we
buffer update row vectors $I$;

\item a counter $t$ of the number of performed {\tt Update} operations
since the last {\tt Init}, modulo $n^\epsilon$.

\end{itemize}

Before proposing our implementation of the operations introduced in
Definition~\ref{def:subquad-matrix}, we discuss a simple invariant
property that we maintain in our data structure and that guarantees
the correctness of the implementation of the operations that we are
going to present.  We use the following notation:

\begin{definition}
We denote by $Buf_J\langle j \rangle$ the $n\times j$ matrix obtained
by considering only the first $j$ columns of $Buf_J$.  Similarly, we
denote by $Buf_I\langle i \rangle$ the $i\times n$ matrix obtained by
considering only the first $i$ rows of $Buf_I$.
\end{definition}

\begin{invariant}
\label{inv:lazy-to-correct}
At any time $t$ in the sequence of operations $\sigma$, the following
invariant is maintained:
$$
M=Lazy+Buf_J\langle t \rangle\cdot Buf_I\langle t \rangle.
$$
\end{invariant}

\bigskip \noindent
\begin{minipage}{15cm}

\noindent{\large \tt Update}~\hrulefill~

\medskip

\begin{frameprog}{15cm}{\small}
\PROCEDURE\ {\tt Update}$(J,I)$ \\
\N \BEGIN \\
\N \> $t\leftarrow t+1$ \\
\N \> \IF\ $t\le n^{\epsilon}$ \THEN \\
\N \> \> $Buf_{J}[\cdot,t]\leftarrow J$ \\
\N \> \> $Buf_{I}[t,\cdot]\leftarrow I$ \\
\N \> \ELSE \\
\N \> \> $t\leftarrow 0$ \\
\N \> \> $Lazy\leftarrow Lazy+Buf_J\cdot Buf_I$ \\
\N \END \\
\end{frameprog}
\end{minipage}

\medskip

\noindent {\tt Update} first increases $t$ and, if $t\le n^{\epsilon}$, it
copies column vector $J$ onto the $t$-th column of $Buf_{J}$ (line
$4$) and row vector $I$ onto the $t$-th row of $Buf_{I}$ (line $5$).
If $t>n^{\epsilon}$, there is no more room in $Buf_{J}$ and $Buf_{I}$
for buffering updates.  Then the counter $t$ is reset in line $7$ and
the reconstruction operation in line $8$ synchronizes $Lazy$ with $M$
via rectangular matrix multiplication of the $n\times n^{\epsilon}$
matrix $Buf_J$ by the $n^{\epsilon}\times n$ matrix $Buf_I$.

\medskip

\bigskip \noindent
\begin{minipage}{15cm}

\noindent{\large \tt Lookup}~\hrulefill~

\medskip

\begin{frameprog}{15cm}{\small}
\PROCEDURE\ {\tt Lookup}$(x,y)$ \\
\N \BEGIN \\
\N \> \RETURN\ $Lazy[x,y]+\sum_{j=1}^{t}Buf_{J}[x,j]\cdot Buf_{I}[j,y]$ \\
\N \END \\
\end{frameprog}
\end{minipage}

\medskip

\noindent {\tt Lookup} runs through the first $t$ columns and rows of 
buffers $Buf_{J}$ and $Buf_{I}$, respectively, and returns the value 
of $Lazy$ corrected with the inner product of the $x$-th row of 
$Buf_{J}\langle t\rangle$ by the $y$-th column of $Buf_{I}\langle 
t\rangle$.

\bigskip \noindent
\begin{minipage}{15cm}

\noindent{\large \tt Init}~\hrulefill~

\medskip

\begin{frameprog}{15cm}{\small}
\PROCEDURE\ {\tt Init}$(X)$ \\
\N \BEGIN \\
\N \> $Lazy\leftarrow X$ \\
\N \> $t\leftarrow 0$ \\
\N \END \\
\end{frameprog}

\medskip

{\tt Init} simply sets the value of $Lazy$ and empties the buffers by
resetting $t$.

\end{minipage}

\bigskip

The following theorem discusses the time and space requirements of 
operations {\tt Update}, {\tt Lookup}, and {\tt Init}.  As already 
stated, the correctness easily follows from the fact that 
Invariant~\ref{inv:lazy-to-correct} is maintained throughout any 
sequence of operations.

\begin{theorem}
\label{th:integer-time}
Each {\tt Update} operation can be supported in
$O(n^{\omega(1,\epsilon,1)-\epsilon})$ worst-case time and each {\tt
Lookup} in $O(n^\epsilon)$ worst-case time, where $0\leq\epsilon\leq
1$ and $\omega(1,\epsilon,1)$ is the exponent for rectangular matrix
multiplication.  {\tt Init} requires $O(n^2)$ time in the worst case.
The space required is $O(n^2)$.
\end{theorem}

\begin{proof}
An amortized update bound follows trivially from amortizing the cost
of the rectangular matrix multiplication $Buf_J\cdot Buf_I$ against
$n^\epsilon$ update operations.  This bound can be made worst-case by
standard techniques, i.e., by keeping two copies of the data
structures: one is used for queries and the other is updated by
performing matrix multiplication in the background.

As fas as {\tt Lookup} is concerned, it answers queries on the value 
of $M[x,y]$ in $\Theta(t)$ worst-case time, where $t\le n^{\epsilon}$.
\end{proof}

\begin{corollary}
\label{co:tc-subq-dynmatr}
If $O(n^\omega)$ is the time required for multiplying two $n\times
n$ matrices, then we can support {\tt Update} in
$O(n^{2-(3-\omega)\epsilon})$ worst-case time and {\tt Lookup} in
$O(n^{\epsilon})$ worst-case time.  Choosing $\epsilon=1$, the
best known bound for matrix multiplication ($\omega<2.38$)
implies an $O(n^{1.38})$ {\tt Update} time and an $O(n)$ {\tt
Lookup} time.
\end{corollary}

\begin{proof}
A rectangular matrix multiplication between a $n\times
n^{\epsilon}$ matrix by a $n^{\epsilon}\times n$ matrix can be
performed by computing $O((n^{1-\epsilon})^2)$ multiplications
between $n^\epsilon \times n^\epsilon$ matrices.  This is done in
$O\left((n^{1-\epsilon})^2\cdot (n^{\epsilon})^\omega\right)$.
The amortized time of the reconstruction operation $Lazy\leftarrow
Lazy+Buf_J\cdot Buf_I$ is thus $O\left(\frac{(n^{1-\epsilon})^2
\cdot
(n^{\epsilon})^\omega+n^2}{n^\epsilon}\right)=O(n^{2-(3-\omega)\epsilon})$.
 The rest of the claim follows from Theorem~\ref{th:integer-time}.
\end{proof}

\section{Transitive Closure Updates in $O({n}^2\log {n})$ Time}
\label{se:tc-log}

In this section we show a first method for casting fully dynamic
transitive closure into the problem of reevaluating polynomials over
Boolean matrices presented in Section~\ref{se:tc-polynomials}.

Based on the technique developed in Section~\ref{se:tc-polynomials}, 
we revisit the dynamic graph algorithm given in~\cite{K99} in terms of 
dynamic matrices and we present a matrix-based variant of it which 
features better initialization time while maintaining the same bounds 
on the running time of update and query operations, i.e., 
$O(n^2\cdot\log n)$ time per update and $O(1)$ time per query.  The 
space requirement of our algorithm is $M(n)\cdot\log n$, where $M(n)$ 
is the space used for representing a polynomial over Boolean matrices.  
As stated in Theorem~\ref{th:polynomials2-complexity}, $M(n)$ is 
$O(n^{2})$ if $h$ and $k$ are constant.

In the remainder of this section we first describe our data structure
and then we show how to support efficiently operations introduced in
Definition~\ref{def:fdkc} for the equivalent problem of fully dynamic
Boolean matrix closure.

\subsection{Data Structure}
\label{ss:tc-log-data-structure}

As it is well known, the Kleene closure of a Boolean matrix $X$ can be 
computed from scratch via matrix multiplication by computing $\log_{2} 
n$ polynomials $P_{k}=P_{k-1}+P_{k-1}^2$, $1\leq k\leq\log_{2}n$.  In 
the static case where $X^*$ has to be computed only once, intermediate 
results can be thrown away as only the final value $X^*=P_{\log_{2} 
n}$ is required.  In the dynamic case, instead, intermediate results 
provide useful information for updating efficiently $X^*$ whenever $X$ 
gets modified.

In this section we consider a slightly different definition of
polynomials $P_1,\ldots,P_{\log_2 n}$ with the property that each of
them has degree $\le 3$:

\begin{definition}
\label{def:dyn_log_decomp}
Let $X$ be an $n\times n$ Boolean matrix.  We define the sequence of
$\log_2 n+1$ polynomials over Boolean matrices $P_0,\ldots,P_{\log_2
n}$ as: $$P_k= \left\{
\begin{array}{ll}
    X & if~k=0 \\
    P_{k-1}+P_{k-1}^2+P_{k-1}^3 & if~k>0  \\
\end{array}
\right.$$
\end{definition}

Before describing our data structure for maintaining the Kleene 
closure of $X$, we discuss some useful properties.

\begin{lemma}
\label{le:dyn_log_decomp_1}
Let $X$ be an $n\times n$ Boolean matrix and let $P_k$ be formed as in
Definition~\ref{def:dyn_log_decomp}.  Then for any $1\leq u,v\leq n$,
$P_k[u,v]=1$ if and only if there is a path $u\leadsto v$ of length at
most $3^k$ in $X$.
\end{lemma}

\begin{proof}
The proof is by induction on $k$.  The base ($k=0$) is trivial.  We 
assume by induction that the claim is satisfied for $P_{k-1}$ and we 
prove that it is satisfied for $P_{k}$ as well.

{\em Sufficient condition}: Any path of length up to $3^k$ between $u$ 
and $v$ in $X$ is either of length up to $3^{k-1}$ or it can be 
obtained as concatenation of three paths of length up to $3^{k-1}$ in 
$X$.  Since all these paths are correctly reported in $P_{k-1}$ by the 
inductive hypothesis, it follows that $P_{k-1}[u,v]=1$ or 
$P_{k-1}^2[u,v]=1$ or $P_{k-1}^3[u,v]=1$.  Thus 
$P_k[u,v]=P_{k-1}[u,v]+P_{k-1}^2[u,v]+P_{k-1}^3[u,v]=1$.

{\em Necessary condition}: If $P_k[u,v]=1$ then at least one among 
$P_{k-1}[u,v]$, $P_{k-1}^2[u,v]$ and $P_{k-1}^3[u,v]$ is $1$.  If 
$P_{k-1}[u,v]=1$, then by the inductive hypothesis there is a path of 
length up to $3^{k-1}<3^k$.  If $P_{k-1}^2[u,v]=1$, then there are two 
paths of length up to $3^{k-1}$ whose concatenation yields a path no 
longer than $3^k$.  Finally, if $P_{k-1}^3[u,v]=1$, then there are 
three paths of length up to $3^{k-1}$ whose concatenation yields a 
path no longer than $3^k$.
\end{proof}

\begin{lemma}
\label{le:dyn_log_decomp_2}
Let $X$ be an $n\times n$ Boolean matrix and let $P_{k}$ be formed as
in Definition~\ref{def:dyn_log_decomp}.  Then $X^*=I_{n}+P_{\log_2
n}$.
\end{lemma}

\begin{proof}
The proof easily follows from Lemma~\ref{le:dyn_log_decomp_1} and from
the observation that that the length of the longest simple path in $X$
is no longer than $n-1<3^{\log_3 n}\le 3^{\log_2 n}$. $I_{n}$ is
required to guarantee the reflexivity of $X^*$.
\end{proof}

Our data structure for maintaining $X^*$ is the following:

\begin{datastructure}
\label{ds:tc-log}
We maintain an $n\times n$ Boolean matrix $X$ and we maintain the 
$\log_{2} n$ polynomials $P_1\ldots P_{\log_{2} n}$ of degree $3$ 
given in Definition~\ref{def:dyn_log_decomp} with instances of Data 
Structure~\ref{ds:poly-hi-degr} presented in 
Section~\ref{se:tc-polynomials}.
\end{datastructure}

As we will see in Section~\ref{ss:tc-log-implementation}, the reason 
for considering the extra term $P_{k-1}^3$ in our data structure is 
that polynomials need to be maintained using not only {\tt 
SetRow}/{\tt SetCol}, but also {\tt LazySet}.  As stated in 
Definition~\ref{def:polynomials}, using {\tt LazySet} yields a weaker 
representation of polynomials, and this forces us to increase the 
degree if complete information about $X^*$ has to be maintained.  This 
aspect will be discussed in more depth in the proof of 
Theorem~\ref{th:log_correct}.

\subsection{Implementation of Operations}
\label{ss:tc-log-implementation}

In this section we show that operations {\tt Init}$^*$, {\tt Set}$^*$,
{\tt Reset}$^*$ and {\tt Lookup}$^*$ introduced in
Definition~\ref{def:fdkc} can all be implemented in terms of
operations {\tt Init}, {\tt LazySet}, {\tt SetRow}, and {\tt SetCol}
(described in Section~\ref{se:tc-polynomials}) on polynomials
$P_1\ldots P_{\log_{2} n}$.

\bigskip \noindent
\begin{minipage}{15cm}

\noindent{\large \tt Init}$^*$~\hrulefill~

\medskip

\begin{frameprog}{15cm}{\small}
\PROCEDURE\ {\tt Init}$^*(X)$ \\
\N \BEGIN \\
\N \> $Y\leftarrow X$ \\
\N \> \FOR\ $k=1$ \TO\ $\log_{2} n$ \DO \\
\N \> \> {\tt P}$_k$.{\tt Init}$(Y)$ \\
\N \> \> $Y\leftarrow ${\tt P}$_k$.{\tt Lookup}$()$ \\
\N \END \\
\end{frameprog}
\end{minipage}

\medskip

\noindent {\tt Init}$^*$ performs {\tt P}$_k$.{\tt Init} operations on 
each $P_{k}$ by propagating intermediate results from $X$ to $P_1$, 
then from $P_1$ to $P_2$, and so on up to $P_{\log_{2} n}$.

\medskip

\bigskip \noindent
\begin{minipage}{15cm}

\noindent{\large \tt Lookup}$^*$~\hrulefill~

\medskip

\begin{frameprog}{15cm}{\small}
\PROCEDURE\ {\tt Lookup}$^*(x,y)$ \\
\N \BEGIN \\
\N \> $Y\leftarrow${\tt P}$_{\log_{2} n}$.{\tt Lookup}$()$ \\
\N \> \RETURN\ $I_{n}+Y[x,y]$ \\
\N \END \\
\end{frameprog}

\medskip

{\tt Lookup}$^*$ returns the value of $P_{\log_{2} n}[x,y]$.

\end{minipage}

\medskip

\bigskip \noindent
\begin{minipage}{15cm}

\noindent{\large \tt Set}$^*$~\hrulefill~

\medskip

\begin{frameprog}{15cm}{\small}
\PROCEDURE\ {\tt Set}$^*(i,\Delta X)$ \\
\N \BEGIN \\
\N \> $\Delta Y\leftarrow \Delta X$ \\
\N \> \FOR\ $k=1$ \TO\ $\log_{2} n$ \DO \\
\N \> \> {\tt P}$_k$.{\tt LazySet}$(\Delta Y,P_{k-1})$ \\
\N \> \> {\tt P}$_k$.{\tt SetRow}$(i,\Delta Y,P_{k-1})$ \\
\N \> \> {\tt P}$_k$.{\tt SetCol}$(i,\Delta Y,P_{k-1})$ \\
\N \> \> $\Delta Y\leftarrow ${\tt P}$_k$.{\tt Lookup}$()$ \\
\N \END \\
\end{frameprog}
\end{minipage}

\medskip

\noindent {\tt Set}$^*$ propagates changes of $P_{k-1}$ to $P_k$ for 
any $k=1$ to $\log_{2} n$.  Notice that any new $1$'s that appear in 
$P_{k-1}$ are inserted in the object $P_{k}$ via {\tt LazySet}, but 
only the $i$-th row and the $i$-th row column of $P_{k-1}$ are taken 
into account by {\tt SetRow} and {\tt SetCol} in order to determine 
changes of $P_k$.  As re-inserting $1$'s already present in a variable 
is allowed by our operations on polynomials, for the sake of 
simplicity in line $7$ we assign the update matrix $\Delta Y$ with 
$P_k$ and not with the variation of $P_k$.

\medskip

\bigskip \noindent
\begin{minipage}{15cm}

\noindent{\large \tt Reset}$^*$~\hrulefill~

\medskip

\begin{frameprog}{15cm}{\small}
\PROCEDURE\ {\tt Reset}$^*(\Delta X)$ \\
\N \BEGIN \\
\N \> $\Delta Y\leftarrow \Delta X$ \\
\N \> \FOR\ $k=1$ \TO\ $\log_{2} n$ \DO \\
\N \> \> $Y\leftarrow ${\tt P}$_k$.{\tt Lookup}$()$ \\
\N \> \> {\tt P}$_k$.{\tt Reset}$(\Delta Y,P_{k-1})$ \\
\N \> \> $\Delta Y\leftarrow Y-${\tt P}$_k$.{\tt Lookup}$()$ \\
\N \END \\
\end{frameprog}
\end{minipage}

\medskip

\noindent {\tt Reset}$^*$ performs {\tt P}$_k$.{\tt Reset} operations 
on each $P_{k}$ by propagating changes specified by $\Delta X$ to 
$P_1$, then changes of $P_1$ to $P_2$, and so on up to $P_{\log_{2} 
n}$.  Notice that we use an auxiliary matrix $Y$ to compute the 
difference between the value of $P_{k}$ before and after the update 
and that the computation of $\Delta Y$ in line $6$ always yields a 
Boolean matrix.

\medskip

\subsection{Analysis}
\label{ss:tc-log-analysis}

In what follows we discuss the correctness and the complexity of our 
implementation of operations {\tt Init}$^*$, {\tt Set}$^*$, {\tt 
Reset}$^*$, and {\tt Lookup}$^*$ presented in 
Section~\ref{ss:tc-log-implementation}.  We recall that $X$ is an 
$n\times n$ Boolean matrix and $P_k$, $0\leq k\leq\log_{2}n$, are the 
polynomials introduced in Definition~\ref{def:dyn_log_decomp}.

\begin{theorem}
\label{th:log_correct}
If at any time during a sequence $\sigma$ of operations there is a
path of length up to $2^k$ between $x$ and $y$ in $X$, then
$P_k[x,y]=1$.
\end{theorem}

\begin{proof}
By induction.  The base is trivial.  We assume that the claim holds 
inductively for $P_{k-1}$, and we show that, after any operation, the 
claim holds also for $P_{k}$.

\begin{itemize}

\item {\tt Init}$^*$: since any {\tt Init}$^*$ operation rebuilds from
scratch $P_{k}$, the claim holds from Lemma~\ref{le:dyn_log_decomp_1}.

\item {\tt Set}$^*$: let us assume that a {\tt Set}$^*$ operation is
performed on the $i$-th row and column of $X$ and a new path $\pi$ of
length up to $2^k$, say $\pi=\langle x, \ldots, i, \ldots, y\rangle$,
appears in $X$ due to this operation.  We prove that $P_k[x,y]=1$
after the operation.

Observe that {\tt P}$_k${\tt.LazySet}$(\Delta P_{k-1},P_{k-1})$ puts 
in place any new $1$'s in any occurrence of the variable $P_{k-1}$ in 
data structure {\tt P}$_{k}$.  We remark that, although the maintained 
value of $P_k$ in data structure {\tt P}$_{k}$ is not updated by {\tt 
LazySet} and therefore the correctness of the current operation is not 
affected, this step is very important: indeed, new $1$'s corresponding 
to new paths of length up to $2^{k-1}$ that appear in $X$ will be 
useful in future {\tt Set}$^*$ operations for detecting the appearance 
of new paths of length up to $2^k$.

If both the portions $x\leadsto i$ and $i \leadsto y$ of $\pi$ have
length up to $2^{k-1}$, then $\pi$ gets recorded in $P_{k-1}^2$, and
therefore in $P_k$, thanks to one of {\tt P$_k$.SetRow}$(i,\Delta
P_{k-1},P_{k-1})$ or {\tt P$_k$.SetCol}$(i,\Delta P_{k-1},P_{k-1})$.
On the other hand, if $i$ is close to (but does not coincide with) one
endpoint of $\pi$, the appearance of $\pi$ may be recorded in
$P_{k-1}^3$, but not in $P_{k-1}^2$.  This is the reason why degree
$2$ does not suffice for $P_k$ in this dynamic setting.

\item {\tt Reset}$^*$: by inductive hypothesis, we assume that
$P_{k-1}[x,y]$ flips to zero after a {\tt Reset}$^*$ operation only if
no path of length up to $2^{k-1}$ remains in $X$ between $x$ and $y$.
Since any {\tt P$_k$.Reset} operation on $P_k$ leaves it as if cleared
$1$'s in $P_{k-1}$ were never set to $1$, $P_{k}[x,y]$ flips to zero
only if no path of length up to $2^k$ remains in $X$.

\end{itemize}
\end{proof}

We remark that the condition stated in Theorem~\ref{th:log_correct} is 
only sufficient because $P_k$ may keep track of paths having length 
strictly more than $2^k$, though no longer than $3^k$.  However, for 
$k=\log_2 n$ the condition is also necessary as no shortest path can 
be longer than $n=2^k$.  Thus, it is straightforward to see that a 
path of any length between $x$ and $y$ exists at any time in $X$ if 
and only if $P_{\log_2 n}[x,y]=1$.

\medskip

The following theorem establishes the running time and space 
requirements of operations {\tt Init}$^*$, {\tt Set}$^*$ and {\tt 
Reset}$^*$.

\begin{theorem}
Any {\tt Init}$^*$ operation can be performed in $O(n^\omega\cdot \log
n)$ worst-case time, where $\omega$ is the exponent of matrix
multiplication; any {\tt Set}$^*$ takes $O(n^2\cdot \log n)$ amortized
time.  The cost of {\tt Reset}$^*$ operations can be charged to
previous {\tt Init}$^*$ and {\tt Set}$^*$ operations.  The maximum
cost charged to each {\tt Init} is $O(n^3\cdot \log n)$.  The space
required is $O(n^2\cdot \log n)$.
\end{theorem}

\begin{proof}
The proof follows from Theorem~\ref{th:polynomials2-complexity} by
considering the time bounds of operations on polynomials described in
Section~\ref{se:tc-polynomials}.  As each maintained polynomial has
constant degree $k=3$, it follows that the space used is $O(n^2\cdot
\log n)$.
\end{proof}

\begin{corollary}
If we perform just one {\tt Init}$^*$ operation in a sequence $\sigma$
of length $\Omega(n)$, or more generally one {\tt Init} operation
every $\Omega(n)$ {\tt Reset} operations, then the amortized cost of
{\tt Reset} is $O(n^2\cdot \log n)$ per operation.
\end{corollary}

\begin{corollary}
If we perform just one {\tt Init}$^*$ operation in a sequence $\sigma$
of length $\Omega(n^2)$, or more generally one {\tt Init} operation
every $\Omega(n^2)$ {\tt Reset} operations, and we perform no
operations {\tt SetRow} and {\tt SetCol}, then the amortized cost of
{\tt Reset} is $O(n\cdot \log n)$ per operation.
\end{corollary}

In the traditional case where {\tt Op}$_{1}=${\tt Init}$^*$ and {\tt
Op}$_{i}\neq${\tt Init}$^*$ for any $i>1$, i.e., {\tt Init}$^*$ is
just performed once at the beginning of the sequence of operations,
previous corollaries state that both {\tt Set}$^*$ and {\tt Reset}$^*$
are supported in $O(n^2\cdot \log n)$ amortized time.  In the
decremental case where only {\tt Reset}$^*$ operations are performed,
the amortized time is $O(n\cdot \log n)$ per update.

\gimmebreak

The algorithm that we presented in this section can be viewed as a 
variant which features very different data structures of the fully 
dynamic transitive closure algorithm presented by King in~\cite{K99}.

King's algorithm is based on a data structure for a graph $G=(V,E)$ 
that maintains a logarithmic number of edge subsets 
$E_0,\ldots,E_{\log_{2} n}$ with the property that $E_{0}=E$ and 
$(x,y)\in E_{i}$ if there is a path $x\leadsto y$ of length up to 
$2^i$ in $G$.  Moreover, if $y$ is not reachable from $x$ in $G$, then 
$(x,y)\not\in E_{i}$ for all $0\le i\le \log_{2} n$.

The maintained values of our polynomials $P_0,\ldots,P_{log_{2} n}$ 
here correspond to the sets $E_{0},\ldots,E_{\log_{2} n}$.

The algorithm by King also maintains $\log_{2} n$ forests 
$F_0,\ldots,F_{\log_{2} n-1}$ such that $F_i$ uses edges in $E_i$ and 
includes $2n$ trees $Out_{i}(v)$ and $In_{i}(v)$, two for each node 
$v\in V$, such that $Out_{i}(v)$ contains all nodes reachable from $v$ 
using at most $2$ edges in $E_i$, and $In_{i}(v)$ contains all nodes 
that reach $v$ using at most $2$ edges in $E_{i}$.  For each pair of 
nodes, also a table $Count_{i}$ is maintained, where $Count_{i}[x,y]$ 
is the number of nodes $v$ such that $x\in In_{i}(v)$ and $y\in 
Out_{i}(v)$.  Now, $E_{i}$ is maintained so as to contain edges 
$(x,y)$ such that $Count_{i-1}[x,y]>0$.  Trees $In_{i}(v)$ and 
$Out_{i}(v)$ are maintained for any node $v$ by means of 
deletions-only data structures~\cite{ES81} which are rebuilt from 
scratch after each $v$-centered insertion of edges.

Our data structures for polynomials over Boolean matrices $P_i$ play 
the same role as King's forests $F_{i}$ of $In_{i}$ and $Out_{i}$ 
trees and of counters $Count_{i}$.

While King's data structures require $O(n^3\cdot \log n)$ worst-case 
initialization time on dense graphs, the strong algebraic properties 
of Boolean matrices allow us to exploit fast matrix multiplication 
subroutines for initializing more efficiently our data structures in 
$O(n^\omega\cdot \log n)$ time in the worst case, where $\omega=2.38$.

\section{Transitive Closure Updates in $O({n}^2)$ Time}
\label{se:tc-divcon}

In this section we show our second and more powerful method for
casting fully dynamic transitive closure into the problem of
reevaluating polynomials over Boolean matrices presented in
Section~\ref{se:tc-polynomials}.

This method hinges upon the well-known equivalence between transitive 
closure and matrix multiplication on a closed semiring and yields a 
new deterministic algorithm that improves the best known bounds for 
fully dynamic transitive closure.  Our algorithm supports each update 
operation in $O(n^2)$ amortized time and answers each reachability 
query with just one matrix lookup.  The space used is $O(n^2)$.

\subsection{Data Structure}
\label{ss:tc-divcon-data-structure}

Let $X$ be a Boolean matrix and let $X^*$ be its Kleene closure.
Before discussing the dynamic case, we recall the main ideas behind
the algorithm for computing statically $X^*$.

\begin{definition}
\label{def:munro_decomp}
Let ${\cal B}_n$ be the set of $n\times n$ Boolean matrices and let
$X\in {\cal B}_n$.  Without loss of generality, we assume that $n$ is
a power of 2.  Define a mapping ${\cal F}:{\cal B}_n\rightarrow
{\cal B}_n$ by means of the following equations:

\vspace{-7mm}
\begin{center}
\begin{equation}
\left\{\begin{tabular}{l}
$E = (A+BD^*C)^{*}$  \\
$F = EBD^*$          \\
$G = D^*CE$          \\
$H = D^*+D^*CEBD^*$  \\
\end{tabular}
\right.
\label{eq:munro_decomp1}
\end{equation}
\end{center}

\noindent where $A, B, C, D$ and $E, F, G, H$ are obtained by
partitioning $X$ and $Y={\cal F}(X)$ into sub-matrices of dimension
$\frac{n}{2}\times\frac{n}{2}$ as follows:

$$X=
\begin{tabular}{|c|c|} \hline
A & B \\ \hline
C & D \\ \hline
\end{tabular} \hspace{10mm}
Y=
\begin{tabular}{|c|c|} \hline
E & F \\ \hline
G & H \\ \hline
\end{tabular}
$$

\end{definition}

\noindent The following fact is well known~\cite{M71}: if $X$ is an 
$n\times n$ Boolean matrix, then ${\cal F}(X)=X^*$.

Another equivalent approach is given below:

\begin{definition}
\label{le:munro_decomp2}
Let ${\cal B}_n$ be the set of $n\times n$ Boolean matrices, let $X\in
{\cal B}_n$ and let ${\cal G}:{\cal B}_n\rightarrow {\cal B}_n$ be the
mapping defined by means of the following equations:

\vspace{-6mm}
\begin{center}
\begin{equation}
\left\{\begin{tabular}{l}
$E = A^*+A^*BHCA^*$  \\
$F = A^*BH$          \\
$G = HCA^*$          \\
$H = (D+CA^*B)^{*}$  \\
\end{tabular}
\right.
\label{eq:munro_decomp2}
\end{equation}
\end{center}

\noindent where $X$ and $Y={\cal G}(X)$ are defined as:
$$X=
\begin{tabular}{|c|c|} \hline
A & B \\ \hline
C & D \\ \hline
\end{tabular} \hspace{10mm}
Y=
\begin{tabular}{|c|c|} \hline
E & F \\ \hline
G & H \\ \hline
\end{tabular}
$$
\end{definition}

It is easy to show that, for any $X\in {\cal B}_n$, ${\cal G}(X)={\cal
F}(X)$. Both ${\cal F}(X)$ and ${\cal G}(X)$ can be computed in
$O(n^\omega)$ worst-case time~\cite{M71}, where $\omega$ is the
exponent of Boolean matrix multiplication.

\medskip

We now define another function ${\cal H}$ such that ${\cal H}(X)=X^*$, 
based on a new set of equations obtained by combining 
Equation~(\ref{eq:munro_decomp1}) and 
Equation~(\ref{eq:munro_decomp2}).  Our goal is to define ${\cal H}$ 
is such a way that it is well-suited for efficient reevaluation in a 
fully dynamic setting.

\begin{lemma}
\label{le:tc-decomp3}
Let ${\cal B}_n$ be the set of $n\times n$ Boolean matrices, let $X\in
{\cal B}_n$ and let ${\cal H}:{\cal B}_n\rightarrow {\cal B}_n$ be the
mapping defined by means of the following equations:

\vspace{-6mm}
\begin{center}
\begin{equation}
\left\{\begin{tabular}{l l l}
$P = D^*$                 & \\
$E_1 = (A+BP^{2}C)^{*}$   & $E_2=E_1BH_2^2CE_1$  & $E=E_1+E_2$  \\
$F_1 = E_1^2BP$           & $F_2=E_1BH_2^2$      & $F=F_1+F_2$  \\
$G_1 = PCE_1^2$           & $G_2=H_2^2CE_1$      & $G=G_1+G_2$  \\
$H_1 = PCE_1^2BP$         & $H_2=(D+CE_1^2B)^*$  & $H=H_1+H_2$  \\
\end{tabular}
\right.
\label{eq:tc-decomp3}
\end{equation}
\end{center}

\noindent where $X$ and $Y={\cal H}(X)$ are defined as:
$$X=
\begin{tabular}{|c|c|} \hline
A & B \\ \hline
C & D \\ \hline
\end{tabular} \hspace{10mm}
Y=
\begin{tabular}{|c|c|} \hline
E & F \\ \hline
G & H \\ \hline
\end{tabular}
$$ Then, for any $X\in {\cal B}_n$, ${\cal H}(X)=X^*.$
\end{lemma}

\begin{proof}
We prove that $E_1+E_2$, $F_1+F_2$, $G_1+G_2$ and $H_1+H_2$ are
sub-matrices of $X^*$:
$$
X^*= \begin{tabular}{|c|c|} \hline
$E_1+E_2$ & $F_1+F_2$ \\ \hline
$G_1+G_2$ & $H_1+H_2$ \\ \hline
\end{tabular}
$$

\medskip

We first observe that, by definition of Kleene closure, 
$X=X^*~\Rightarrow~X=X^2$.  Thus, since $E_1 = (A+BP^{2}C)^{*}$, 
$H_2=(D+CE_1^2B)^*$ and $P=D^*$ are all closures, then we can replace 
$E_1^2$ with $E_1$, $H_2^2$ with $H_2$ and $P^2$ with $P$.  This 
implies that $E_1=(A+BPC)^{*}=(A+BD^*C)^{*}$ and then $E_1=E$ by 
Equation~\ref{eq:munro_decomp1}.  Now, $E$ is a sub-matrix of $X^*$ 
and encodes explicitly all paths in $X$ with both end-points in 
$V_1=\{1,\ldots,\frac{n}{2}\}$, and since $E_2=EB(D+CEB)^*CE$, then 
$E_2\subseteq E$.  It follows that $E_1+E_2=E+E_2=E$.  With a similar 
argument, we can prove that $F_1+F_2$, $G_1+G_2$ and $H_1+H_2$ are 
sub-matrices of $X^*$.  In particular, for $H=H_1+H_2$ we also need to 
observe that $D^*\subseteq H_2$.
\end{proof}

Note that ${\cal H}$ provides a method for computing the Kleene 
closure of an $n\times n$ Boolean matrix, provided that we are able to 
compute Kleene closures of Boolean matrices of size $\frac{n}{2}\times 
\frac{n}{2}$.  The reason of using $E_{1}^2$, $H_{2}^2$ and $P^2$ 
instead of $E_1$, $H_{2}$ and $P$ in Equation~(\ref{eq:tc-decomp3}), 
which is apparently useless, will be clear in 
Lemma~\ref{le:tc-det-square} after presenting a fully dynamic version 
of the algorithm that defines ${\cal H}$.

In the next lemma we show that a Divide and Conquer algorithm that 
recursively uses ${\cal H}$ to solve sub-problems of smaller size 
requires asymptotically the same time as computing the product of two 
Boolean matrices.

\begin{theorem}
\label{th:h-runtime}
Let $X$ be an $n\times n$ Boolean matrix and let $T(n)$ be the time
required to compute recursively ${\cal H}(X)$.  Then
$T(n)=O(n^\omega)$, where $O(n^\omega)$ is the time required to
multiply two Boolean matrices.
\end{theorem}

\begin{proof}
It is possible to compute $E$, $F$, $G$ and $H$ with three recursive
calls of ${\cal H}$, a constant number $c_m$ of multiplications, and a
constant number $c_s$ of additions of $\frac{n}{2}\times\frac{n}{2}$
matrices.  Thus: $$T(n)\le 3\,T(\frac{n}{2})+c_m
M(\frac{n}{2})+c_s\left(\frac{n}{2}\right)^2$$
where $M(n)=O(n^\omega)$ is the time required to multiply two $n\times 
n$ Boolean matrices.  Solving the recurrence relation, since 
$\log_{2}3<\max\{\omega,2\}=\omega$, we obtain that $T(n)=O(n^\omega)$ 
(see e.g., the Master Theorem in~\cite{CLR90}).
\end{proof}

The previous theorem showed that, even if ${\cal H}$ needs to compute
one more closure than ${\cal F}$ and ${\cal G}$, asymptotically the
running time does not get worse.

\gimmebreak

In the following, we study how to reevaluate efficiently ${\cal 
H}(X)=X^*$ under changes of $X$.  Our data structure for maintaining 
the Kleene closure $X^*$ is the following:

\begin{datastructure}
\label{ds:tc-det}
We maintain two $n\times n$ Boolean matrices $X$ and $Y$ decomposed in
sub-matrices $A$, $B$, $C$, $D$, and $E$, $F$, $G$, $H$:
$$
X=
\begin{tabular}{|c|c|} \hline
A & B \\ \hline
C & D \\ \hline
\end{tabular} \hspace{10mm}
Y=
\begin{tabular}{|c|c|} \hline
E & F \\ \hline
G & H \\ \hline
\end{tabular}
$$
We also maintain the following $12$ polynomials over $n\times n$
Boolean matrices with the data structure presented in
Section~\ref{se:tc-polynomials}:
$$
\begin{tabular}{l l l}
$Q = A+BP^{2}C$           & $E_2=E_1BH_2^2CE_1$  & $E=E_1+E_2$  \\
$F_1 = E_1^2BP$           & $F_2=E_1BH_2^2$      & $F=F_1+F_2$  \\
$G_1 = PCE_1^2$           & $G_2=H_2^2CE_1$      & $G=G_1+G_2$  \\
$H_1 = PCE_1^2BP$         & $R=D+CE_1^2B$        & $H=H_1+H_2$  \\
\end{tabular}
$$
and we recursively maintain $3$ Kleene closures $P$, $E_1$ and $H_2$:
$$
\begin{tabular}{l l l}
$P = D^*$ & $E_1 = Q^*$ & $H_2=R^*$  \\
\end{tabular}
$$
with instances of size $\frac{n}{2}\times\frac{n}{2}$ of Data 
Structure~\ref{ds:poly-hi-degr} presented in 
Section~\ref{se:tc-polynomials}
\end{datastructure}

\begin{figure}
\centerline{ { \epsfxsize=13cm \epsffile{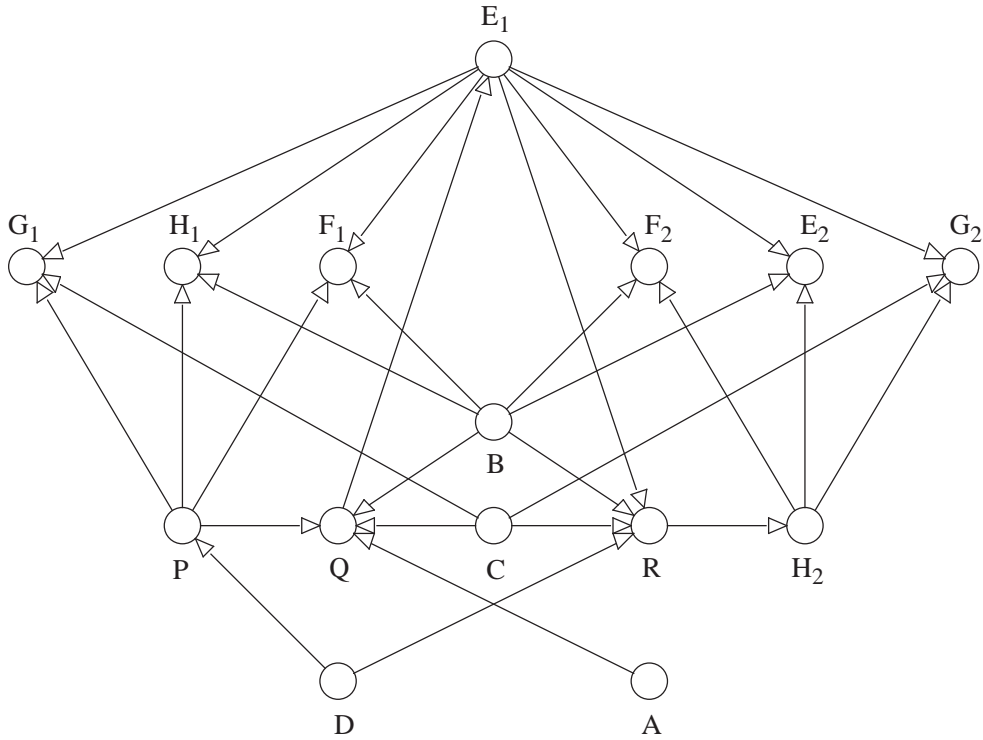}} }
\caption{Data dependencies between polynomials and closures.}
\label{fi:tc-det-acycdep}
\end{figure}

It is worth to note that Data Structure~\ref{ds:tc-det} is recursively
defined: $P$, $E_1$ and $H_2$ are Kleene closures of
$\frac{n}{2}\times\frac{n}{2}$ matrices.  Also observe that the
polynomials $Q$, $F_1$, $G_1$, $H_1$, $E_2$, $F_2$, $G_2$, $R$, $E$,
$F$, $G$ and $H$ that we maintain have all constant degree $\le 6$.
In Figure~\ref{fi:tc-det-acycdep} we show the acyclic graph of
dependencies between objects in our data structure: there is an arc
from node $u$ to node $v$ if the polynomial associated to $u$ is a
variable of the polynomial associated to $v$.  For readability, we do
not report nodes for the final polynomials $E$, $F$, $G$, $H$.  A
topological sort of this graph, e.g., $\tau=\langle P$, $Q$, $E_1$,
$R$, $H_2$, $F_1$, $G_1$, $H_1$, $E_2$, $F_2$, $G_2$, $E$, $F$, $G$,
$H\rangle$, yields a correct evaluation order for the objects in the
data structure and thus gives a method for computing ${\cal H}(X)$.

We remark that our data structure has memory of all the intermediate 
values produced when computing ${\cal H}(X)$ from scratch and 
maintains such values upon updates of $X$.  As it was already observed 
in Section~\ref{se:tc-log}, maintaining intermediate results of some 
static algorithm for computing $X^*$ is a fundamental idea for 
updating efficiently $X^*$ whenever $X$ gets modified.  

Since our data structure reflects the way ${\cal H}(X)$ is computed,
it basically represents $X^*$ as the sum of two Boolean matrices: the
first, say $X_1^*$, is defined by submatrices $E_1, F_1, G_1, H_1$,
and the second, say $X_2^*$, by submatrices $E_2, F_2, G_2, H_2$:
$$
X_{1}^*= \begin{tabular}{|c|c|} \hline
$E_1$ & $F_1$ \\ \hline
$G_1$ & $H_1$ \\ \hline
\end{tabular} \hspace{10mm}
X_{2}^*=
\begin{tabular}{|c|c|} \hline
$E_2$ & $F_2$ \\ \hline
$G_2$ & $H_2$ \\ \hline
\end{tabular}
$$

In the next section we show how to implement operations {\tt
Init}$^*$, {\tt Set}$^*$, {\tt Reset}$^*$ and {\tt Lookup}$^*$
introduced in Definition~\ref{def:fdkc} in terms of operations {\tt
Init}, {\tt LazySet}, {\tt SetRow} and {\tt SetCol} (see
Section~\ref{se:tc-polynomials}) on the polynomials of Data
Structure~\ref{ds:tc-det}.

\subsection{Implementation of Operations}
\label{ss:tc-divcon-implementation}

\begin{figure}[t]
\centerline{ { \epsfxsize=13.5cm \epsffile{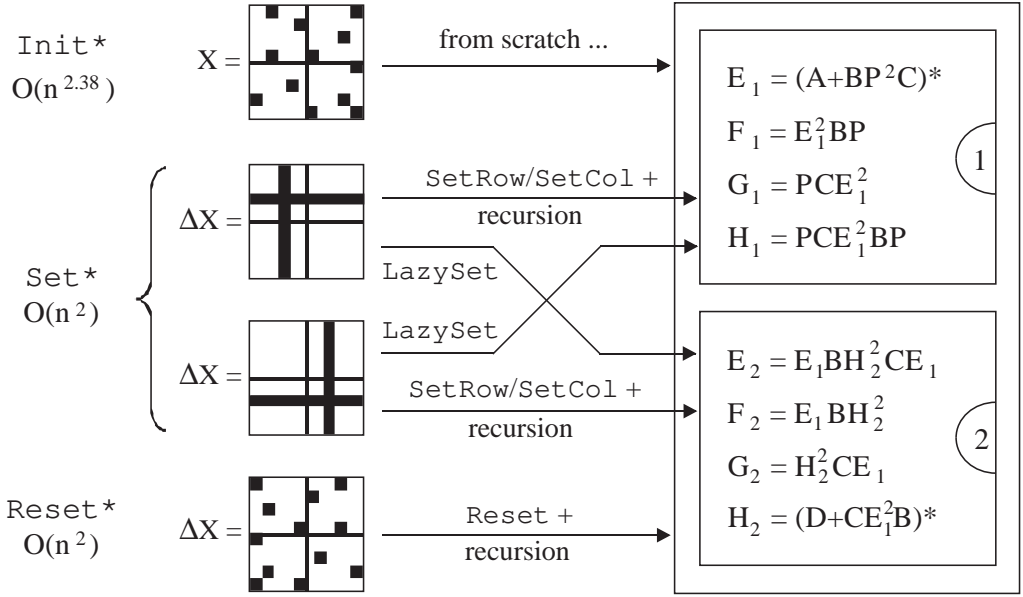}} }
\caption{Overview of operations {\tt Init}$^*$, {\tt Set}$^*$ and  {\tt
Reset}$^*$.}
\label{fi:tc-det-operations}
\end{figure}

From a high-level point of view, our approach is the following.  We 
maintain $X_1^*$ and $X_2^*$ {\em in tandem} (see 
Figure~\ref{fi:tc-det-operations}): whenever a {\tt Set}$^*$ operation 
is performed on $X$, we update $X^*$ by computing how either $X_1^*$ 
or $X_2^*$ are affected by this change.  Such updates are lazily 
performed so that neither $X_1^*$ nor $X_2^*$ encode complete 
information about $X^*$, but their sum does.  On the other side, {\tt 
Reset}$^*$ operations update both $X_1^*$ and $X_2^*$ and leave the 
data structures as if any reset entry was never set to $1$.

We now describe in detail our implementation.  To keep pseudocodes 
shorter and more readable, we assume that implicit {\tt Lookup} and 
{\tt Lookup}$^*$ operations are performed in order to retrieve the 
current value of objects so as to use them in subsequent steps.  
Furthermore, we do not deal explicitly with base recursion steps.

\bigskip \noindent
\begin{minipage}{15cm}

\noindent{\large \tt Init}$^*$~\hrulefill~

\medskip

\begin{frameprog}{15cm}{\small}
\PROCEDURE\ {\tt Init}$^*(Z)$ \\
\N \BEGIN \\
\N \> $X\leftarrow Z$ \\
\N \> {\tt P.Init}$^*(D)$   \\
\N \> {\tt Q.Init}$(A,B,P,C)$   \\
\N \> {\tt E$_1$.Init}$^*(Q)$   \\
\N \> {\tt R.Init}$(D,C,E_1,B)$ \\
\N \> {\tt H$_2$.Init}$^*(R)$   \\
\N \> {\tt F$_1$.Init}$(E_1,B,P)$ \\
\N \> \{ similarly for $G_1$, $H_1$, $E_2$, $F_2$, $G_2$, and
then for $E$, $F$, $G$, $H$ \} \\
\N \END \\
\end{frameprog}
\end{minipage}

\medskip

\noindent{\tt Init}$^*$ sets the initial value of $X$ (line $2$) and
initializes the objects in Data Structure~\ref{ds:tc-det} according to
the topological order $\tau$ of the graph of dependencies as explained
in the previous subsection (lines $3$--$9$).

\medskip

\medskip\noindent\begin{minipage}{15cm}

\noindent{\large \tt Set}$^*$~\hrulefill~

\medskip Before describing our implementation of {\tt Set}$^*$, we
first define a useful shortcut for performing simultaneous {\tt
SetRow} and {\tt SetCol} operations with the same $i$ on more than one
variable in a polynomial $P$:
\end{minipage}

\medskip\noindent\begin{minipage}{15cm}
\begin{frameprog}{15cm}{\small}
\PROCEDURE\ {\tt P.Set}$(i,\Delta X_1,\ldots,\Delta X_q)$    \\
\N \BEGIN \\
\N \> {\tt P.SetRow}$(i,\Delta X_1,X_1)$ \\
\N \> {\tt P.SetCol}$(i,\Delta X_1,X_1)$ \\
\N \> \> \> ~~\vdots\\
\N \> {\tt P.SetRow}$(i,\Delta X_q,X_q)$ \\
\N \> {\tt P.SetCol}$(i,\Delta X_q,X_q)$ \\
\N \END \\
\end{frameprog}
\end{minipage}

\medskip \noindent Similarly, we give a shortcut\footnote{For the sake
of simplicity, we use the same identifier {\tt LazySet} for both the
shortcut and the native operation on polynomials, assuming to use the
shortcut in defining {\tt Set}$^*$.} for performing simultaneous {\tt
LazySet} operations on more than one variable in a polynomial $P$:

\medskip\noindent\begin{minipage}{15cm}
\begin{frameprog}{15cm}{\small}
\PROCEDURE\ {\tt P.LazySet}$(\Delta X_1,\ldots,\Delta X_q)$    \\
\N \BEGIN \\
\N \> {\tt P.LazySet}$(\Delta X_1,X_1)$ \\
\N \> \> \> ~~\vdots\\
\N \> {\tt P.LazySet}$(\Delta X_q,X_q)$ \\
\N \END \\
\end{frameprog}
\end{minipage}

\medskip\noindent\begin{minipage}{15cm} We also define an auxiliary 
operation {\tt LazySet}$^*$ on closures that performs {\tt LazySet} 
operations for variables $A$, $B$, $C$ and $D$ on the polynomials $Q$, 
$R$, $F_1$, $G_1$, $H_1$, $E_2$, $F_2$, and $G_2$ and recurses on the 
closure $P$ which depend directly on them.  We assume that, if $M$ is 
a variable of a polynomial maintained in our data structure, $\Delta 
M=M_{curr}-M_{old}$ is the difference between the current value 
$M_{curr}$ of $M$ and the old value $M_{old}$ of $M$.
\end{minipage}

\medskip\noindent\begin{minipage}{15cm}
\begin{frameprog}{15cm}{\small}
\PROCEDURE\ {\tt LazySet}$^*(\Delta X)$    \\
\N \BEGIN \\
\N \> $X\leftarrow X+\Delta X$ \\
\N \> {\tt Q.LazySet}$(\Delta A,\Delta B,\Delta C)$ \\
\N \> {\tt R.LazySet}$(\Delta B,\Delta C,\Delta D)$ \\
\N \> \{ similarly for $F_1$, $G_1$, $H_1$, $E_2$, $F_2$, and $G_2$ \} \\
\N \> {\tt P.LazySet}$^*(\Delta D)$ \\
\N \END \\
\end{frameprog}
\end{minipage}

\medskip\noindent\begin{minipage}{15cm} Using the shortcuts {\tt Set}
and {\tt LazySet} and the new operation {\tt LazySet}$^*$, we are now
ready to define {\tt Set}$^*$.
\end{minipage}

\medskip\noindent\begin{minipage}{15cm}
\begin{frameprog}{15cm}{\small}
\PROCEDURE\ {\tt Set}$^*(i,\Delta X)$                         \\
\N \BEGIN                                                     \\
\N \> $X\leftarrow X+I_{\Delta X,i}+J_{\Delta X,i}$           \\
\N \> \IF\ $1\le i\le \frac{n}{2}$ \THEN                      \\
\N \> \> {\tt Q.Set}$(i,\Delta A,\Delta B,\Delta C)$          \\
\N \> \> {\tt E$_1$.Set}$^*(i,\Delta Q)$                      \\
\N \> \> {\tt F$_1$.Set}$(i,\Delta E_1,\Delta B)$             \\
\N \> \> {\tt G$_1$.Set}$(i,\Delta C,\Delta E_1)$             \\
\N \> \> {\tt H$_1$.Set}$(i,\Delta C,\Delta E_1,\Delta B)$    \\
\N \> \> {\tt R.Set}$(i,\Delta C,\Delta E_1,\Delta B)$        \\
\N \> \> {\tt H$_2$.LazySet}$^*(\Delta R)$                    \\
\N \> \> {\tt G$_2$.LazySet}$(\Delta C,\Delta E_1)$           \\
\N \> \> {\tt F$_2$.LazySet}$(\Delta E_1,\Delta B)$           \\
\N \> \> {\tt E$_2$.LazySet}$(\Delta E_1,\Delta B,\Delta C)$  \\
\N \> \ELSE\ \{ $\frac{n}{2}+1\le i\le n$ \}                  \\
\N \> \> $i\leftarrow i-\frac{n}{2}$                          \\
\N \> \> {\tt P.Set}$^*(i,\Delta D)$                          \\
\N \> \> {\tt R.Set}$(i,\Delta B,\Delta C,\Delta D)$          \\
\N \> \> {\tt H$_2$.Set}$^*(i,\Delta R)$                      \\
\N \> \> {\tt G$_2$.Set}$(i,\Delta H_2,\Delta C)$             \\
\N \> \> {\tt F$_2$.Set}$(i,\Delta B,\Delta H_2)$             \\
\N \> \> {\tt E$_2$.Set}$(i,\Delta B,\Delta H_2,\Delta C)$    \\
\N \> \> {\tt Q.Set}$(i,\Delta B,\Delta P,\Delta C)$          \\
\N \> \> {\tt E$_1$.LazySet}$^*(\Delta Q)$                    \\
\N \> \> {\tt F$_1$.LazySet}$(\Delta B,\Delta P)$             \\
\N \> \> {\tt G$_1$.LazySet}$(\Delta P,\Delta C)$             \\
\N \> \> {\tt H$_1$.LazySet}$(\Delta B,\Delta P,\Delta C)$    \\
\N \> {\tt E.Init}$(E_1,E_2)$                                 \\
\N \> {\tt F.Init}$(F_1,F_2)$                                 \\
\N \> {\tt G.Init}$(G_1,G_2)$                                 \\
\N \> {\tt H.Init}$(H_1,H_2)$                                 \\
\N \END \\
\end{frameprog}
\end{minipage}

\medskip

\noindent {\tt Set}$^*$ performs an $i$-centered update in $X$ and
runs through the closures and the polynomials of Data
Structure~\ref{ds:tc-det} to propagate any changes of $A$, $B$, $C$,
$D$ to $E$, $F$, $G$, $H$.  The propagation order is $\langle Q$,
$E_1$, $F_1$, $G_1$, $H_1$, $R$, $H_2$, $G_2$, $F_2$, $E_2$, $E$, $F$,
$G$, $H\rangle$ if $1\le i\le \frac{n}{2}$ and $\langle P$, $R$,
$H_2$, $G_2$, $F_2$, $E_2$, $Q$, $E_1$, $F_1$, $G_1$, $H_1\rangle$ if
$\frac{n}{2}+1\le i\le n$ and is defined according to a topological
sort of the graph of dependencies between objects in Data
Structure~\ref{ds:tc-det} shown in Figure~\ref{fi:tc-det-acycdep}.

Roughly speaking, {\tt Set}$^*$ updates the objects in the data
structure according to the value of $i$ as follows:

\begin{enumerate}

\item If $1\le i\le \frac{n}{2}$, fully updates $Q$, $R$, $E_1$,
$F_1$, $G_1$, $H_1$ (lines $4$--$9$) and lazily updates $E_2$, $F_2$,
$G_2$, $H_2$ (lines $10$--$13$).  See
Figure~\ref{fi:tc-det-acycdep-set} (a).

\begin{figure}[p]
\centerline{ { \epsfysize=20cm \epsffile{tc-det-acycdep-set.epsf}} }
\caption{Portions of Data Structure~\ref{ds:tc-det} affected during a
{\tt Set}$^*$ operation when: (a) $1\le i\le \frac{n}{2}$; (b)
$\frac{n}{2}+1\le i\le n$ .}
\label{fi:tc-det-acycdep-set}
\end{figure}

\item If $\frac{n}{2}+1\le i\le n$, fully updates $P$, $Q$, $R$,
$E_2$, $F_2$, $G_2$, $H_2$ (lines $16$--$22$) and lazily updates
$E_1$, $F_1$, $G_1$, $H_1$ (lines $23$--$26$).  See
Figure~\ref{fi:tc-det-acycdep-set} (b).

\end{enumerate}

We highlight that it is not always possible to perform efficiently 
full updates of all the objects of Data Structure~\ref{ds:tc-det}.  
Actually, some objects may change everywhere, and not only in a row 
and column.  Such unstructured changes imply that we can only perform 
lazy updates on such objects, as they cannot be efficiently 
manipulated by means of $i$-centered {\tt SetRow} and {\tt SetCol} 
operations.

We now explain in detail the operations performed by {\tt Set}$^*$
according to the two cases $1\le i\le \frac{n}{2}$ and
$\frac{n}{2}+1\le i\le n$.

\subsubsection*{Case 1: $1\le i\le \frac{n}{2}$.}

In this case an $i$-centered update of $X$ may affect the $i$-th row
and the $i$-th column of $A$, the $i$-th row of $B$ and the $i$-th
column of $C$, while $D$ is not affected at all by this kind of update
(see Figure~\ref{fi:tc-det-operations}).  The operations performed by
{\tt Set}$^*$ when $1\le i\le \frac{n}{2}$ are therefore the
following:

\begin{description}

\item[Line $2$:] an $i$-centered set operation is performed on $X$.

\item[Line $4$:]
$Q=A+BP^{2}C$ is updated by performing {\tt SetRow} and {\tt SetCol} 
operations for any variables $A$, $B$ and $C$ being changed.  
$P=D^{*}$ does not change since, as already observed, $D$ is not 
affected by the change.  Notice that new $1$'s may appear in $Q$ only 
in the $i$-th row and column due to this operation.

\item[Line $5$:]
{\tt Set}$^*$ is recursively called to propagate the changes of $Q$ to 
$E_1$.  We remark that $E_1$ may change also outside the $i$-th row 
and column due to this operation.  Nevertheless, as we will see in 
Lemma~\ref{le:tc-det-closure}, the fact that $E_1$ is a closure 
implies that new $1$'s appear in a very structured way.  This will 
make it possible to propagate changes efficiently to any polynomial 
that, in turn, depends on $E_1$.

\item[Lines $6$--$9$:]
polynomials $F_1$, $G_1$, $H_1$ and $R$ are updated by performing {\tt
SetRow} and {\tt SetCol} operations for any variables $E_1$, $B$ and
$C$ being changed.  We recall that such operations take into account
only the entries of $\Delta E_1$ lying in the $i$-th row and in the
$i$-th column, albeit other entries may be non-zero.  Again,
Lemma~\ref{le:tc-det-closure} and Lemma~\ref{le:tc-det-square} will
show that this is sufficient.

\item[Lines $10$--$13$:]
$H_2=R^*$ is not updated, but new $1$'s that appear in $R$ are lazily 
inserted in the data structure of $H_2$ by calling {\tt LazySet}$^*$.  
Then {\tt LazySet} operations are carried out on polynomials $G_2$, 
$F_2$, $E_2$ to insert in the data structures that maintain them any 
new $1$'s that appear in $C$, $E_1$ and $B$.

\item[Lines $27$--$30$.]
Recompute polynomials $E$, $F$, $G$ and $H$ from scratch.  This is 
required as $F_1$, $G_1$ and $H_2$ may change everywhere and not only 
in a row and a column.  Differently from the case of $E_1$, whose 
change is structured as it is a closure, we cannot exploit any 
particular structure of $\Delta F_1$, $\Delta G_1$ and $\Delta H_2$ 
for reducing ourselves to use {\tt SetRow} and {\tt SetCol} and we are 
forced to use {\tt Init}.  Note that, since $E$, $F$, $G$ and $H$ have 
all degree $1$, this is not a bottleneck in terms of running time.

\end{description}

\subsubsection*{Case 2: $\frac{n}{2}+1\le i\le n$.}

In this case an $i$-centered update of $X$ may affect only the $i$-th 
row and the $i$-th column of $D$, the $i$-th row of $C$ and the $i$-th 
column of $B$, while $A$ is not affected at all by this kind of update 
(see Figure~\ref{fi:tc-det-operations}).

Operations performed by {\tt Set}$^*$ are completely analogous to the 
case $1\le i\le \frac{n}{2}$, except for the fact that we need to 
rescale the index $i$ in line $15$ and we have also to perform a 
recursive call to update $P$ in line $16$.

\medskip

\bigskip \noindent
\begin{minipage}{15cm}

\noindent{\large \tt Reset}$^*$~\hrulefill~
\end{minipage}

\medskip

\noindent Before describing our implementation of {\tt Reset}$^*$, we
define a useful shortcut\footnote{For the sake of simplicity, we use
the same identifier {\tt Reset} for both the shortcut and the native
operation on polynomials, assuming to use the shortcut in defining
{\tt Reset}$^*$.} for performing simultaneous {\tt Reset} operations
on more than one variable in a polynomial $P$.

\medskip\noindent\begin{minipage}{15cm}
\begin{frameprog}{15cm}{\small}
\PROCEDURE\ {\tt P.Reset}$(\Delta X_1,\ldots,\Delta X_q)$    \\
\N \BEGIN \\
\N \> {\tt P.Reset}$(\Delta X_1,X_1)$ \\
\N \> \> \> ~~\vdots\\
\N \> {\tt P.Reset}$(\Delta X_q,X_q)$ \\
\N \END \\
\end{frameprog}
\end{minipage}

\medskip

\noindent Using this shortcut, we are now ready to define {\tt
Reset}$^*$.  We assume that, if $M$ is a variable of a polynomial
maintained in our data structure, $\Delta M=M_{old}-M_{curr}$ is the
difference between the value $M_{old}$ of $M$ just before calling {\tt
Reset}$^*$ and the current value $M_{curr}$ of $M$.

\medskip\noindent\begin{minipage}{15cm}
\begin{frameprog}{15cm}{\small}
\PROCEDURE\ {\tt Reset}$^*(\Delta X)$ \\
\N \BEGIN \\
\N \> $X\leftarrow X-\Delta X$ \\
\N \> {\tt P.Reset}$^*(\Delta D)$   \\
\N \> {\tt Q.Reset}$(\Delta A,\Delta B,\Delta P,\Delta C)$   \\
\N \> {\tt E$_1$.Reset}$^*(\Delta Q)$   \\
\N \> {\tt R.Reset}$(\Delta D,\Delta C,\Delta E_1,\Delta B)$ \\
\N \> {\tt H$_2$.Reset}$^*(\Delta R)$   \\
\N \> {\tt F$_1$.Reset}$(\Delta E_1,\Delta B,\Delta P)$ \\
\N \> \{ similarly for $G_1$, $H_1$, $E_2$, $F_2$, $G_2$, and
then for $E$, $F$, $G$, $H$ \} \\
\N \END \\
\end{frameprog}
\end{minipage}

\medskip

\noindent {\tt Reset}$^*$ resets any entries of $X$ as specified by 
$\Delta X$ and runs through the closures and the polynomials in the 
data structure to propagate any changes of $A$, $B$, $C$, $D$ to $E$, 
$F$, $G$, $H$.  The propagation is done according to a topological 
order $\tau$ of the graph of dependencies shown in 
Figure~\ref{fi:tc-det-acycdep} and is the same order followed by {\tt 
Init}$^*$, which has a similar structure.  Actually, we could think of 
{\tt Reset}$^*$ as a function that ``undoes'' any previous work 
performed by {\tt Init}$^*$ and {\tt Set}$^*$ on the data structure, 
leaving it as if the reset entries of $X$ were never set to $1$.

\medskip

\bigskip \noindent
\begin{minipage}{15cm}

\noindent{\large \tt Lookup}$^*$~\hrulefill~
\end{minipage}

\medskip\noindent\begin{minipage}{15cm}
\begin{frameprog}{15cm}{\small}
\PROCEDURE\ {\tt Lookup}$^*(x,y)$    \\
\N \BEGIN \\
\N \> \RETURN\ $Y[x,y]$ \\
\N \END \\
\end{frameprog}

\medskip

{\tt Lookup}$^*$ simply returns the maintained value of $Y[x,y]$.

\end{minipage}

\subsection{Analysis}
\label{ss:tc-divcon-analysis}

Now we discuss the correctness and the complexity of our 
implementation.  Before providing the main claims, we give some 
preliminary definitions and lemmas that are useful for capturing 
algebraic properties of the changes that polynomials in our data 
structure undergo during a {\tt Set}$^*$ operation.

The next definition recalls a property of Boolean update matrices that 
is related to the operational concept of $i$-centered update.

\begin{definition}
\label{def:i-centered}
We say that a Boolean update matrix $\Delta X$ is $i-centered$ if
$\Delta X=I_{\Delta X,i}+J_{\Delta X,i}$, i.e., all entries lying
outside the $i$-th row and the $i$-th column are zero.
\end{definition}

If the variation $\Delta X$ of some matrix $X$ during an update 
operation is $i$-centered and $X$ is a variable of a polynomial $P$ 
that has to be efficiently reevaluated, then we can use {\tt P.SetRow} 
and {\tt P.SetCol} operations which are especially designed for doing 
so.  But what happens if $X$ changes by a $\Delta X$ that is not 
$i$-centered?  Can we still update efficiently the polynomial $P$ 
without recomputing it from scratch via {\tt Init}?  This is the case 
of $E_1$ and $\Delta E_1$ while performing a {\tt Set}$^*$ update with 
$1\le i\le \frac{n}{2}$.  In the following we show that, under certain 
hypotheses on $X$ and $\Delta X$ (which are satisfied by $E_1$ and 
$\Delta E_1$), we can still solve the problem efficiently.

While the property of being $i$-centered is related to an update
matrix by itself, the following two definitions are concerned with
properties of an update matrix $\Delta X$ with respect to the matrix
$X$ to which it is applied:

\begin{definition}
\label{def:i-transitive}
If $X$ is a Boolean matrix and $\Delta X$ is a Boolean update matrix, 
we say that $\Delta X$ is $i$-transitive with respect to $X$ if 
$I_{\Delta X,i}=I_{\Delta X,i}\cdot X$ and $J_{\Delta X,i}=X\cdot 
J_{\Delta X,i}$.
\end{definition}

\begin{definition}
\label{def:i-complete}
If $X$ is a Boolean matrix and $\Delta X$ is a Boolean update matrix, 
we say that $\Delta X$ is $i$-complete with respect to $X$ if $\Delta 
X=J_{\Delta X,i}\cdot I_{\Delta X,i}+X\cdot I_{\Delta X,i}+J_{\Delta 
X,i}\cdot X$.
\end{definition}

Using the previous definitions we can show that the variation of
$X^*$ due to an $i$-centered update of $X$ is $i$-transitive and
$i$-complete.

\begin{lemma}
\label{le:tc-det-closure}
Let $X$ be a Boolean matrix and let $\Delta X$ be an $i$-centered 
update matrix.  If we denote by $\Delta X^*$ the matrix $(X+\Delta 
X)^*-X^*$, then $\Delta X^*$ is $i$-transitive and $i$-complete with 
respect to $X^*$.
\end{lemma}

\begin{proof}
The following equalities prove the first condition of 
$i$-transitivity: $$I_{\Delta X^*,i}\cdot X^* = I_{(X+\Delta 
X)^*-X^*,i}\cdot X^* = I_{(X+\Delta X)^*\cdot X^* - X^*\cdot X^*,i} = 
I_{(X+\Delta X)^*-X^*,i} = I_{\Delta X^*,i}.$$ 

The other conditions can be proved analogously.  The hypothesis that 
$\Delta X$ is $i$-centered is necessary for the $i$-completeness.
\end{proof}

The following lemma shows under what conditions for $\Delta X$ and $X$ 
it is possible to perform operations of the kind $X\leftarrow X+\Delta 
X$ on a variable $X$ of a polynomial by reducing such operations to 
$i$-centered updates even if $\Delta X$ is not $i$-centered.

\begin{lemma}
\label{le:tc-det-square}
If $X$ is a Boolean matrix such that $X=X^*$ and $\Delta X$ is an
$i$-transitive and $i$-complete update matrix with respect to $X$,
then $X+\Delta X=(X+I_{\Delta X,i}+J_{\Delta X,i})^2$.
\end{lemma}

\begin{proof}
Since $X=X^*$ it holds that $X=X^2$ and $X=X+I_{\Delta X,i}\cdot 
J_{\Delta X,i}$.  The proof follows from 
Definition~\ref{def:i-transitive} and Definition~\ref{def:i-complete} 
and from the facts that: $I_{\Delta X,i}^2\subseteq I_{\Delta X,i}$, 
$J_{\Delta X,i}^2\subseteq J_{\Delta X,i}$ and $\Delta X=\Delta 
X+I_{\Delta X,i}+J_{\Delta X,i}$.
\end{proof}

It follows that, under the hypotheses of Lemma~\ref{le:tc-det-square},
if we replace any occurrence of $X$ in $P$ with $X^2$ and we perform
both {\tt P.SetRow}$(i,I_{\Delta X,i},X)$ and {\tt
P.SetCol}$(i,J_{\Delta X,i},X)$, then new $1$'s in $P$ correctly
appear.  This is the reason why in Data Structure~\ref{ds:tc-det} we
used $E_{1}^2$, $H_{2}^2$, and $P^2$ instead of $E_1$, $H_{2}$, and
$P$, respectively.

Before stating the main theorem of this section which establishes the 
correctness of operations on our data structure, we discuss a general 
property of polynomials and closures over Boolean matrices that will 
be useful in proving the theorem.

\begin{lemma}
\label{le:poly-composition}
Let $P$ and $Q$ be polynomials or closures over Boolean matrices and
let $\widehat{P}$ and $\widehat{Q}$ be relaxed functions such that
$\widehat{P}(X)\subseteq P(X)$ and $\widehat{Q}(Y)\subseteq Q(Y)$ for any
values of variables $X$ and $Y$.  Then, for any $X$:
$$\widehat{Q}(\widehat{P}(X))\subseteq Q(P(X))$$
\end{lemma}

\begin{proof}
Let $\widehat{Y}=\widehat{P}(X)$ and $Y=P(X)$.  By definition, we 
have: $\widehat{Y}\subseteq Y$ and $\widehat{Q}(\widehat{Y})\subseteq 
Q(\widehat{Y})$.  By exploiting a monotonic behavior of polynomials 
and closures over Boolean matrices, we have: $\widehat{Y}\subseteq 
Y\Rightarrow Q(\widehat{Y})\subseteq Q(Y)$.  Thus: 
$\widehat{Q}(\widehat{Y})\subseteq Q(\widehat{Y})\subseteq 
Q(Y)\Rightarrow\widehat{Q}(\widehat{Y})\subseteq 
Q(Y)\Rightarrow\widehat{Q}(\widehat{P}(X))\subseteq Q(P(X))$.
\end{proof}

\begin{theorem}
Let ${\cal H}$ be the function defined in Lemma~\ref{le:tc-decomp3},
let $X$ and $Y$ be the matrices maintained in Data
Structure~\ref{ds:tc-det}, and let $M$ be a Boolean matrix whose value
at any time $j$ is defined as:
$$
M_j=\hspace{-3mm}\sum_{\scriptsize
\begin{array}{c}
1\le i\le j: \\
{\tt Op}_i\neq {\tt LazySet}^*\\
\end{array}}\hspace{-3mm} {\cal H}(X_{i})-{\cal H}(X_{i-1}).
$$
If we denote by $X_j$ and $Y_j$ the values of $X$ and $Y$ after the
$j$-th operation, respectively, then the relation $M_{j}\subseteq
Y_j\subseteq {\cal H}(X_{j})$ is satisfied.
\end{theorem}

\begin{proof}
The proof is by induction on the size $n$ of matrices in Data 
Structure~\ref{ds:tc-det}.  The base is trivial.  We assume that the 
claim holds for instances of size $\frac{n}{2}$ and we prove that it 
holds also for instances of size $n$.

\begin{itemize}

\item {\tt Op$_{j}$=Init}$^*$: since {\tt Init}$^*$ performs {\tt Init}
operations on each object, then $Y_j={\cal H}(X_j)$.

\item {\tt Op$_{j}$=Set}$^*$: we first prove that $Y_j\subseteq {\cal 
H}(X_{j})$.  Observe that $Y$ is obtained as a result of a composition 
of functions that relax the correct intermediate values of polynomials 
and closures of Boolean matrices in our data structure allowing them 
to contain less $1$'s.  Indeed, by the properties of {\tt Lookup} 
described in Section~\ref{se:tc-polynomials}, we know that, if $P$ is 
the correct value of a polynomial at any time, then {\tt 
P.Lookup}$()\subseteq P$.  Similarly, by inductive hypothesis, if $K$ 
is a Kleene closure of an $\frac{n}{2}\times \frac{n}{2}$ Boolean 
matrix, then at any time {\tt K.Lookup}$^*(x,y)=1\Rightarrow 
K[x,y]=1$.  The claim then follows by Lemma~\ref{le:poly-composition}, 
which states that the composition of relaxed functions computes values 
containing at most the $1$'s contained in the values computed by the 
correct functions.

To prove that $M_{j}\subseteq Y_j$, based on the definition of $M$, it
suffices to verify that $\Delta {\cal H}(X)\subseteq \Delta Y$, where
$\Delta {\cal H}(X)={\cal H}(X_{j})-{\cal H}(X_{j-1})$ and $\Delta
Y=Y_{j}-Y_{j-1}$.  In particular, we prove that if ${\cal H}[x,y]$
flips from $0$ to $1$ due to operation {\tt Set}$^*$, then either
$X_1^*[x,y]$ flips from $0$ to $1$ (due to lines $4$--$8$ when $1\le
i\le \frac{n}{2}$), or $X_2^*[x,y]$ flips from $0$ to $1$ (due to
lines $17$--$21$ when $\frac{n}{2}+1\le i\le n$).

Without loss of generality, assume that the {\tt Set}$^*$ operation is
performed with $1\le i\le \frac{n}{2}$ (the proof is completely
analogous if $\frac{n}{2}+1\le i\le n$).

As shown in Figure~\ref{fi:tc-det-operations}, sub-matrices $A$, $B$
and $C$ may undergo $i$-centered updates due to this operation and so
their variation can be correctly propagated through {\tt SetRow} and
{\tt SetCol} operations to polynomial $Q$ (line $4$) and to
polynomials $F_1$, $G_1$ and $H_1$ (lines $6$--$8$).  As $\Delta Q$ is
also $i$-centered due to line $4$, any variation of $Q$, that is
assumed to be elsewhere correct from previous operations, can be
propagated to closure $E_1$ through a recursive call of {\tt Set$^*$}
in line $5$.  By the inductive hypothesis, this propagation correctly
reveals any new $1$'s in $E_1$.  We remark that $E_1$ may contain less
$1$'s than $E$ due to any previous {\tt LazySet} operations done in
line $23$.

Observe now that $E_1$ occurs in polynomials $F_1$, $G_1$ and $H_1$
and that $\Delta E_1$ is not necessarily $i$-centered.  This would
imply that we cannot propagate directly changes of $E_1$ to these
polynomials, as no efficient operation for doing so was defined in
Section~\ref{se:tc-polynomials}.  However, by
Lemma~\ref{le:tc-det-closure}, $\Delta E_1$ is $i$-transitive and
$i$-complete with respect to $E_1$.  Since $E_1=E_1^*$, by
Lemma~\ref{le:tc-det-square} performing both {\tt SetRow}$(i,I_{\Delta
E_1,i},E_1)$ and {\tt SetCol}$(i,J_{\Delta E_1,i},E_1)$ operations on
data structures {\tt F$_1$}, {\tt G$_1$} and {\tt H$_1$} in lines
$6$--$8$ is sufficient to correctly reveal new $1$'s in $F_1$, $G_1$
and $H_1$.

Again, note that $F_1$, $G_1$ and $H_1$ may contain less $1$'s than
$F$, $G$ and $H$, respectively, due to any previous {\tt LazySet}
operations done in lines $23$--$26$.  We have then proved that lines
$4$--$8$ correctly propagate any $i$-centered update of $X$ to
$X_1^*$.

To conclude the proof, we observe that $E_1$ also occurs in
polynomials $E_2$, $F_2$, $G_2$, $R$ and indirectly affects $H_2$.
Unfortunately, we cannot update $H_2$ efficiently as $\Delta R$ is
neither $i$-centered, nor $i$-transitive/$i$-complete with respect to
$R$.  So in lines $9$--$13$ we limit ourselves to update explicitly $R$ and
to log any changes of $E_1$ by performing {\tt LazySet} operations on
polynomials $G_2$, $F_2$, and $E_2$ and a {\tt LazySet}$^*$ operation on
$H_2$.  This is sufficient to guarantee the correctness of subsequent
{\tt Set}$^*$ operations for $\frac{n}{2}+1\le i\le n$.

\item {\tt Op$_{j}$=Reset}$^*$: this operation runs in judicious
order through the objects in the data structure and undoes the
effects of previous {\tt Set}$^*$ and {\tt Init}$^*$ operations.  Thus,
any property satisfied by $Y$ still holds after performing a {\tt
Reset}$^*$ operation.

\end{itemize}
\end{proof}

\begin{corollary}
Let {\tt X} be an instance of Data Structure~\ref{ds:tc-det} and let
$\sigma=\langle {\tt X.Op}_1,\ldots,$ ${\tt X.Op}_k\rangle$ be a
sequence of operations on {\tt X}.  If ${\tt Op}_i\neq {\tt
LazySet}^*$ for all $1\le i\le j\le k$, then $M_j={\cal H}(X_j)$.
\end{corollary}

\begin{proof}
Since ${\cal H}(0_n)=0_{n}^*=0_{n}$, the proof easily follows by 
telescoping the sum that defines $M_j$: $M_j={\cal H}(X_{j})-{\cal 
H}(X_{j-1})+{\cal H}(X_{j-1})-{\cal H}(X_{j-2})+\cdots+{\cal 
H}(X_2)-{\cal H}(X_1)+{\cal H}(X_1)-{\cal H}(X_0)={\cal H}(X_j)-{\cal 
H}(X_0)={\cal H}(X_j).$
\end{proof}

To conclude this section, we address the running time of operations 
and the space required to maintain an instance of our data structure.

\begin{theorem}
Any {\tt Init}$^*$ operation can be performed in $O(n^\omega)$
worst-case time, where $\omega$ is the exponent of matrix
multiplication; any {\tt Set}$^*$ takes $O(n^2)$ amortized time.  The
cost of {\tt Reset}$^*$ operations can be charged to previous {\tt
Init}$^*$ and {\tt Set}$^*$ operations.  The maximum cost charged to
each {\tt Init}$^*$ is $O(n^3)$.  The space required is $O(n^2)$.
\end{theorem}

\begin{proof}
Since all the polynomials in Data Structure~\ref{ds:tc-det} are of
constant degree and involve a constant number of terms, the amortized
cost of any {\tt SetRow}, {\tt SetCol}, {\tt LazySet}, and {\tt Reset}
operation on them is quadratic in $\frac{n}{2}$ (see
Theorem~\ref{th:polynomials2-complexity}).  Let $T(n)$ be the time
complexity of any {\tt Set}$^*$, {\tt LazySet}$^*$ and {\tt Reset}$^*$
operation.  Then: $$T(n)\leq 3\,T(\frac{n}{2})+\frac{c\,n^2}{4}$$for
some suitably chosen constant $c>0$.  As $\log_{2}3<2$, this implies
that $T(n)=O(n^2)$.

{\tt Init}$^*$ recomputes recursively ${\cal H}$ from scratch using 
{\tt Init} operations on polynomials, which require $O(n^\omega)$ 
worst-case time each.  We can then prove that the running time of {\tt 
Init}$^*$ is $O(n^\omega)$ exactly as in Theorem~\ref{th:h-runtime}.

\medskip

To conclude the proof, observe that if $K(n)$ is the space used to
maintain all the objects in Data Structure~\ref{ds:tc-det}, and $M(n)$
is the space required to maintain a polynomial with the data structure
of Section~\ref{se:tc-polynomials}, then:
$$
K(n)\leq 3\,K(\frac{n}{2})+12\,M(n).
$$
Since $M(n)=O(n^2)$ by Theorem~\ref{th:polynomials2-complexity}, then 
$K(n)=O(n^2)$.
\end{proof}

\begin{corollary}
If we perform just one {\tt Init}$^*$ operation in a sequence $\sigma$ 
of length $\Omega(n)$, or more generally one {\tt Init}$^*$ operation 
every $\Omega(n)$ {\tt Reset}$^*$ operations, then the amortized cost 
of {\tt Reset}$^*$ is $O(n^2)$ per operation.
\end{corollary}

\begin{corollary}
If we perform just one {\tt Init}$^*$ operation in a sequence $\sigma$ 
of length $\Omega(n^2)$, or more generally one {\tt Init}$^*$ 
operation every $\Omega(n^2)$ {\tt Reset}$^*$ operations, and we 
perform no {\tt Set}$^*$ operations, then the amortized cost of {\tt 
Reset}$^*$ is $O(n)$ per operation.
\end{corollary}

In the traditional case where {\tt Op}$_{1}=${\tt Init}$^*$ and {\tt 
Op}$_{i}\neq${\tt Init}$^*$ for any $i>1$, i.e., {\tt Init}$^*$ is 
performed just once at the beginning of the sequence of operations, 
previous corollaries state that both {\tt Set}$^*$ and {\tt Reset}$^*$ 
are supported in $O(n^2)$ amortized time.  In the decremental case 
where only {\tt Reset}$^*$ operations are performed, the amortized 
time is $O(n)$ per update.

\section{Breaking Through the $O(n^{2})$ Barrier}
\label{se:tc-subquad}

In this section we present the first algorithm that supports both 
updates and queries in subquadratic time per operation, showing that 
it is actually possible to break through the $O(n^2)$ barrier on the 
single-operation complexity of fully dynamic transitive closure.  This 
result is obtained by means of a new technique that consists of 
casting fully dynamic transitive closure into the problem of 
dynamically maintaining matrices over integers presented in 
Section~\ref{se:tc-subquadraticmatrices}.  As already shown in 
Section~\ref{se:tc-log} and in Section~\ref{se:tc-divcon}, dynamic 
matrices, thanks to their strong algebraic properties, play a crucial 
role in designing efficient algorithms for the fully dynamic 
transitive closure problem.

The remainder of this section is organized as follows.  In 
Section~\ref{se:tc-subquad-counting} we present a subquadratic 
algorithm for directed acyclic graphs based on dynamic matrices that 
answers queries in $O(n^\epsilon)$ time and performs updates in 
$O(n^{\omega(1,\epsilon,1)-\epsilon}+n^{1+\epsilon})$ time, for any 
$0\le \epsilon\le 1$, where $\omega(1,\epsilon,1)$ is the exponent of 
the multiplication of an $n\times n^{\epsilon}$ matrix by an 
$n^{\epsilon}\times n$ matrix.  According to the current best bounds 
on $\omega(1,\epsilon,1)$, we obtain an $O(n^{0.58})$ query time and 
an $O(n^{1.58})$ update time.  The algorithm we propose is randomized, 
and has one-side error.

\subsection{Counting Paths in Acyclic Directed Graphs}
\label{se:tc-subquad-counting}

In this section we study a variant of the fully dynamic transitive 
closure problem presented in Definition~\ref{def:fdtc} and we devise 
the first algorithm that supports both update and query in 
subquadratic time per operation.  In the variant that we consider, the 
graph that we maintain is constrained to be acyclic; furthermore, {\tt 
Insert} and {\tt Delete} operations work on single edges rather than 
on set of edges.  We shall discuss later how to extend our algorithm 
to deal with more than one edge at a time.

\begin{definition}
\label{def:fdtc-acyc}
Let $G=(V,E)$ be a directed acyclic graph and let $TC(G)=(V,E')$ be
its transitive closure.  We consider the problem of maintaining a data
structure {\tt G} for the graph $G$ under an intermixed sequence
$\sigma=\langle {\tt G.Op}_1,\ldots,{\tt G.Op}_k\rangle$ of update and
query operations.  Each operation ${\tt G.Op}_j$ on the data structure
{\tt G} can be either one of the following:

\begin{itemize}

\item {\tt G.Insert}$(x,y)$: perform the update $E\leftarrow
E\cup\{(x,y)\}$, such that the graph obtained after the update is
still acyclic.

\item {\tt G.Delete}$(x,y)$: perform the update $E\leftarrow
E-\{(x,y)\}$, where $(x,y)\in E$.

\item {\tt G.Query}$(x,y)$: perform a query operation on $TC(G)$ by 
returning $1$ if $(x,y)\in E'$ and $0$ otherwise.

\end{itemize}

\end{definition}

In this version of the problem, we do not deal explicitly with
initialization operations.

\subsubsection*{Data Structure}

In \cite{KS99} King and Sagert showed that keeping a count of the
number of distinct paths between any pair of vertices in a directed
acyclic graph $G$ allows it to maintain the transitive closure of $G$
upon both insertions and deletions of edges.  Unfortunately, these
counters may be as large as $2^n$: to perform $O(1)$ time arithmetic
operations on counters, an $O(n)$ wordsize is required.  As shown in
\cite{KS99}, the wordsize can be reduced to $2c\lg n$ for any $c\geq
5$ based on the use of arithmetic operations performed modulo a random
prime number.  This yields a fully dynamic randomized Monte Carlo
algorithm for transitive closure with the property that ``yes''
answers on reachability queries are always correct, while ``no''
answers are wrong with probability $O(\frac{1}{n^c})$.  We recall that
this algorithm performs reachability queries in $O(1)$ and updates in
$O(n^{2})$ worst-case time on directed acyclic graphs.

We now present an algorithm that combines the path counting approach 
of King and Sagert with our technique of implicit matrix 
representation.  Both techniques are very simple, but surprisingly 
their combination solves a problem that has been open for many years.

\begin{datastructure}
We keep a count of the number of distinct paths between any pair of 
vertices in graph $G$ by means of an instance $M$ of the dynamic 
matrix data structure described in 
Section~\ref{se:tc-subquadraticmatrices}.  We assume that $M[x,y]$ is 
the number of distinct paths between node $x$ and node $y$ in graph 
$G$.  Since $G$ is acyclic, this number is well-defined.
\end{datastructure}

\subsubsection*{Implementation of Operations}

We now show how to implement operations {\tt Insert}, {\tt Delete} and 
{\tt Query} in terms of operations {\tt Update} and {\tt Lookup} on 
our data structure as described in 
Section~\ref{se:tc-subquadraticmatrices}.  We assume all arithmetic 
operations are performed in constant time.

\bigskip \noindent
\begin{minipage}{15cm}

\noindent{\large \tt Insert}~\hrulefill~
\end{minipage}

\medskip\noindent\begin{minipage}{15cm}
\begin{frameprog}{15cm}{\small}
\PROCEDURE\ {\tt Insert}$(x,y)$    \\
\N \BEGIN \\
\N \> $E\leftarrow E\cup\{(x,y)\}$ \\
\N \> \FOR\ $z=1$ \TO\ $n$ \DO \\
\N \> \> $J[z]\leftarrow {\tt M.Lookup}(z,x)$ \\
\N \> \> $I[z]\leftarrow {\tt M.Lookup}(y,z)$ \\
\N \> {\tt M.Update}$(J,I)$ \\
\N \END \\
\end{frameprog}
\end{minipage}

\medskip\noindent {\tt Insert} first puts edge $(x,y)$ in the graph 
and then, after querying matrix {\tt M}, computes two vectors $J$ and 
$I$ such that $J[z]$ is the number of distinct paths $z\leadsto x$ in 
$G$ and $I[z]$ is the number of distinct paths $y\leadsto z$ in $G$ 
(lines $3$--$5$).  Finally, it updates {\tt M} in line $6$.  The 
operation performed on $M$ is $M\leftarrow M+J\cdot I$: this means 
that the number $M[u,v]$ of distinct paths between any two nodes 
$(u,v)$ is increased by the number $J[u]$ of distinct paths $u\leadsto 
x$ times the number $I[v]$ of distinct paths $y\leadsto v$, i.e., 
$M[u,v]\leftarrow M[u,v]+J[u]\cdot I[v]$.

\bigskip \noindent
\begin{minipage}{15cm}

\noindent{\large \tt Delete}~\hrulefill~
%

\begin{frameprog}{15cm}{\small}
\PROCEDURE\ {\tt Delete}$(x,y)$    \\
\N \BEGIN \\
\N \> $E\leftarrow E-\{(x,y)\}$ \\
\N \> \FOR\ $z=1$ \TO\ $n$ \DO \\
\N \> \> $J[z]\leftarrow {\tt M.Lookup}(z,x)$ \\
\N \> \> $I[z]\leftarrow {\tt M.Lookup}(y,z)$ \\
\N \> {\tt M.Update}$(-J,I)$ \\
\N \END \\
\end{frameprog}
\end{minipage}

\medskip \noindent {\tt Delete} is identical to {\tt Insert}, except 
for the fact that it removes the edge $(x,y)$ from the graph and 
performs the update of {\tt M} in line $6$ with $-J$ instead of $J$.  
The operation performed on $M$ is $M\leftarrow M-J\cdot I$: this means 
that the number $M[u,v]$ of distinct paths between any two nodes 
$(u,v)$ is decreased by the number $J[u]$ of distinct paths $u\leadsto 
x$ times the number $I[v]$ of distinct paths $y\leadsto v$, i.e., 
$M[u,v]\leftarrow M[u,v]-J[u]\cdot I[v]$.

\bigskip \noindent
\begin{minipage}{15cm}

\noindent{\large \tt Query}~\hrulefill~
\end{minipage}

\medskip\noindent\begin{minipage}{15cm}
\begin{frameprog}{15cm}{\small}
\PROCEDURE\ {\tt Query}$(x,y)$    \\
\N \BEGIN \\
\N \> \IF\ {\tt M.Lookup}$(x,y)>0$ \THEN\ \RETURN\ 1 \\
\N \> \ELSE\ \RETURN\ 0 \\
\N \END \\
\end{frameprog}
\end{minipage}

\medskip \noindent
{\tt Query} simply looks up the value of $M[x,y]$ and returns $1$ if
the current number of distinct paths between $x$ and $y$ is positive,
and zero otherwise.

\gimmebreak

We are now ready to discuss the running time of our implementation of 
operations {\tt Insert}, {\tt Delete}, and {\tt Query}.

\begin{theorem}
Any {\tt Insert} and any {\tt Delete} operation can be performed in
$O(n^{\omega(1,\epsilon,1)-\epsilon}+n^{1+\epsilon})$ worst-case time,
for any $0\le \epsilon\le 1$, where $\omega(1,\epsilon,1)$ is the
exponent of the multiplication of an $n\times n^{\epsilon}$ matrix by
an $n^{\epsilon}\times n$ matrix.  Any {\tt Query} takes
$O(n^\epsilon)$ in the worst case.  The space required is $O(n^2)$.
\end{theorem}

\begin{proof}
We recall that, by Theorem \ref{th:integer-time}, each entry of $M$
can be queried in $O(n^{\epsilon})$ worst-case time, and each {\tt
Update} operation can be performed in
$O(n^{\omega(1,\epsilon,1)-\epsilon})$ worst-case time.  Since $I$ and
$J$ can be computed in $O(n^{1+\epsilon})$ worst-case time by means of
$n$ queries on $M$, we can support both insertions and deletions in
$O(n^{\omega(1,\epsilon,1)-\epsilon}+n^{1+\epsilon})$ worst-case time,
while a reachability query for any pair of vertices $(x,y)$ can be
answered in $O(n^\epsilon)$ worst-case time by simply querying the
value of $M[x,y]$.
\end{proof}

\begin{corollary}
Any {\tt Insert} and any {\tt Delete} operation requires
$O(n^{1.58})$ worst-case time, and any {\tt Query} requires
$O(n^{0.58})$ worst-case time.
\end{corollary}

\begin{proof}
Balancing the two terms in the update bound
$O(n^{\omega(1,\epsilon,1)-\epsilon}+n^{1+\epsilon})$ yields that
$\epsilon$ must satisfy the equation
$\omega(1,\epsilon,1)=1+2\epsilon$.  The current best bounds on
$\omega(1,\epsilon,1)$~\cite{CW90,HP98} imply that
$\epsilon<0.58$~\cite{Z98}.  Thus, the smallest update time is
$O(n^{1.58})$, which gives a query time of $O(n^{0.58})$.
\end{proof}

The algorithm we presented is deterministic.  However, as the numbers 
involved may be as large as $2^n$, performing arithmetic operations in 
constant time requires wordsize $O(n)$.  To reduce wordsize to $O(\log 
n)$ while maintaining the same subquadratic bounds ($O(n^{1.58})$ per 
update and $O(n^{0.58})$ per query) we perform all arithmetic 
operations modulo some random prime number as explained in 
\cite{KS99}.  Again, this produces a randomized Monte Carlo algorithm, 
where ``yes'' answers on reachability queries are always correct, 
while ``no'' answers are wrong with probability $O(\frac{1}{n^c})$ for 
any constant $c\geq 5$.

It is also not difficult to extend our subquadratic algorithm to deal 
with insertions/deletions of more than one edge at a time.  In 
particular, we can support any insertion/deletion of up to 
$O(n^{1-\eta})$ edges incident to a common vertex in 
$O(n^{\omega(1,\epsilon,1)-\epsilon}+n^{2-(\eta-\epsilon)})$ 
worst-case time.  We emphasize that this is still $o(n^2)$ for any 
$1>\eta>\epsilon>0$.  Indeed, rectangular matrix multiplication can be 
trivially implemented via matrix multiplication: this implies that 
$\omega(1,\epsilon,1)<2-(2-\omega)\epsilon$, where 
$\omega=\omega(1,1,1)<2.38$ is the current best exponent for matrix 
multiplication~\cite{CW90}.

\section{Conclusions}
\label{se:tc-conclusions}

In this paper we have presented new time and space efficient 
algorithms for maintaining the transitive closure of a directed graph 
under edge insertions and edge deletions.  As a main contribution, we 
have introduced a general framework for casting fully dynamic 
transitive closure into the problem of dynamically reevaluating 
polynomials over matrices when updates of variables are performed.  
Such technique has turned out to be very flexible and powerful, 
leading both to revisit the best known algorithm for fully dynamic 
transitive closure~\cite{K99} from a completely different perspective, 
and to design new and faster algorithms for the problem.

In particular, efficient data structures for maintaining polynomials 
over Boolean matrices allowed us to devise the fairly complex 
deterministic algorithm described in Section~\ref{se:tc-divcon}, which 
supports updates in quadratic amortized time and queries with just one 
matrix lookup.  Our algorithm improves the best bounds for fully 
dynamic transitive closure achieved in~\cite{K99} and is the fastest 
algorithm with constant query time known in literature for this 
problem.

In addition, a surprisingly simple technique for efficiently 
maintaining dynamic matrices of integers under simultaneous updates of 
multiple entries, combined with a previous idea of counting paths in 
acyclic digraphs~\cite{KS99}, yielded the randomized algorithm 
presented in Section~\ref{se:tc-subquad-counting}: this algorithm, for 
the first time in the study of fully dynamic transitive closure, 
breaks through the $O(n^{2})$ barrier on the single-operation 
complexity of the problem.

\section*{Acknowledgements}

We are indebted to Garry Sagert and Mikkel Thorup for enlightening 
discussions, and to Valerie King for many useful comments and insights 
on this work.

\bibliography{AlgoBiblio}

\end{document}